\documentclass[final,3p,times,authoryear]{elsarticle}

\usepackage{orcidlink}

\usepackage[utf8]{inputenc}

\usepackage{amsmath}
\usepackage{amsfonts}
\usepackage{booktabs}
\usepackage{multirow}
\usepackage{siunitx} 
\usepackage{adjustbox}
\pdfoutput=1
\usepackage{graphicx}
\usepackage{float}

\usepackage[caption=false]{subfig}

\usepackage{braket}
\usepackage{dcolumn}
\usepackage{bm,url}

\usepackage{placeins}

\usepackage{ulem} 

\usepackage{multirow}
\usepackage{dsfont}

\usepackage{adjustbox}
\newcommand{\be}{\begin{equation}}
\newcommand{\ee}{\end{equation}}
\newcommand{\bea}{\begin{eqnarray}}
\newcommand{\eea}{\end{eqnarray}}

\DeclareMathAlphabet{\mathcal}{OMS}{cmsy}{m}{n}

\usepackage[dvipsnames,svgnames,table]{xcolor}  
\usepackage{ragged2e}

\usepackage{orcidlink}

\definecolor{oxfordblue}{rgb}{0.0, 0.13, 0.28}
\definecolor{burgundy}{rgb}{0.5, 0.0, 0.13}
\definecolor{darkolivegreen}{rgb}{0.33, 0.42, 0.18}
\definecolor{darkblue}{rgb}{0,0,0.5}
\definecolor{richcarmine}{rgb}{0.84, 0.0, 0.25}
\definecolor{darkblue}{rgb}{0,0,0.5}
\definecolor{bluer}{rgb}{0.00,0.50,0.75}{}
\hypersetup{colorlinks=true, citecolor=red, linkcolor=blue,
 urlcolor = magenta, filecolor=magenta}

\usepackage{ragged2e}

\usepackage{amssymb}
\usepackage{lipsum}

\journal{High Energy Astrophysics}

\begin{document}

\begin{frontmatter}


\title{Non-minimally coupled loop quantum inflation with inverse-volume corrections}


\author[tarapaca]{Rudranil Roy\orcidlink{0000-0003-3114-419X}}
\ead{rudranil.roy@alumnos.uta.cl}
\author[tarapaca]{Giovanni Otalora\orcidlink{0000-0001-6753-0565}}
\ead{giovanni.otalora@academicos.uta.cl}
\affiliation[tarapaca]{organization={Departamento de F\'isica, Facultad de Ciencias, Universidad de Tarapac\'a},
            addressline={Casilla 7-D}, 
            city={Arica},
            country={Chile}}

\author[joel]{Joel Saavedra\orcidlink{0000-0002-1430-3008}}
\ead{joel.saavedra@pucv.cl}

\affiliation[joel]{organization={Instituto de F\'{\i}sica, Pontificia Universidad Cat\'olica de Valpara\'{\i}so},
            addressline={Casilla 4950}, 
            city={Valpara\'{\i}so},
            country={Chile}}
\author[salvatore_1,salvatore_2,salvatore_3]{Salvatore Capozziello\orcidlink{0000-0003-4886-2024}}
\ead{capozziello@na.infn.it}

\affiliation[salvatore_1]{organization={Dipartimento di Fisica ``E. Pancini'', Universit\`a degli Studi di Napoli ``Federico II''},
            addressline={Complesso Universitario di Monte Sant'Angelo, Edificio G, Via Cinthia, I-80126}, 
            city={Napoli},
            country={Italy}}

\affiliation[salvatore_2]{organization={Istituto Nazionale di Fisica Nucleare (INFN)},
            addressline={Sezione di Napoli, Via Cinthia 9, I-80126}, 
            city={Napoli},
            country={Italy}}

\affiliation[salvatore_3]{organization={Scuola Superiore Meridionale},
            addressline={Via Mezzocannone 4, I-80134}, 
            city={Napoli},
            country={Italy}}

\begin{abstract}
We study slow-roll inflation driven by a scalar field non-minimally coupled to gravity within the effective framework of Loop Quantum Cosmology (LQC), including inverse-volume corrections. We consider two physically motivated classes of potentials, a Higgs-like quartic potential $V\propto\phi^{4}$ and string-inspired fractional monomial potentials $V\propto\phi^{p}$ with $p<1$. Working at first order in the slow-roll expansion, we derive analytic expressions for the inflationary observables, namely the scalar spectral index $n_s$, the tensor-to-scalar ratio $r$, and the running $\alpha_s\equiv dn_s/d\ln k$, and then solve the corrected background dynamics numerically to obtain quantitative predictions. Confronting these results with current observational constraints from Planck 2018 and ACT DR6, we find that the model can lie within the allowed region of the $(n_s,r,\alpha_s)$ parameter space, including a mild preference for slightly larger $n_s$, as suggested by recent ground-based measurements. We also compute the probability of achieving sufficient slow-roll inflation in this setting. Although effective LQC replaces the initial singularity with a nonsingular quantum bounce, the likelihood of a sufficiently long inflationary phase depends on the pre-inflationary dynamics and the inflaton potential. Using the canonical Liouville measure on the effective phase space, we determine the fraction of post-bounce trajectories that yield sufficient inflation and find that the non-minimal coupling parameter $\xi$ substantially enlarges the phase-space volume of favorable initial conditions relative to the minimally coupled case, exhibiting an attractor-like enhancement that saturates at large $\xi$.
\end{abstract}







\end{frontmatter}




\section{Introduction}
\label{sec:Introduction}

Cosmic inflation has emerged as the leading theoretical framework for describing the earliest stages of the universe \cite{Guth:1980zm,Starobinsky:1980te,Linde:1981mu,Albrecht:1982wi}. It provides a natural explanation for the observed homogeneity, isotropy, and flatness of the cosmos, while also accounting for the primordial fluctuations imprinted in the cosmic microwave background (CMB) and the large-scale structure of the universe \cite{Mukhanov:1981xt,Bardeen:1983qw,Planck:2018jri,Liddle:2000cg}. Numerous inflationary models successfully reproduce current observational constraints, as reviewed in several foundational works \cite{Martin:2013tda,Kallosh:2025ijd,Gao:2025viy,Ellis:2025zrf,Kumar:2025apf}. Despite these successes, key conceptual challenges remain, including the sensitivity of inflation to initial conditions \cite{Goldwirth:1989pr,Brandenberger:2016uzh,Linde:2017pwt,Clough:2016ymm,East:2015ggf} and the need for a consistent ultraviolet (UV) completion that embeds inflation within a fundamental quantum gravitational theory \cite{Burgess:2007pt,Baumann:2014nda}.

Loop Quantum Cosmology (LQC), a symmetry-reduced quantization derived from Loop Quantum Gravity \cite{Ashtekar:1986yd,Ashtekar:1987gu,Rovelli:1987df,Rovelli:1989za,Rovelli:2004tv,Thiemann:2007pyv,Ashtekar:2004eh,Perez:2012wv,Gambini:2011zz}, offers an appealing avenue for addressing these issues. One of its most notable predictions is the replacement of the big-bang singularity by a quantum bounce, a consequence of the discrete quantum geometry underlying the theory \cite{Ashtekar:2011ni,Ashtekar:2006rx}. This modification introduces non-classical phases such as superinflation and alters the effective Friedmann dynamics \cite{Copeland:2007qt,Bojowald:2002nz,Bojowald:2003mc,Ashtekar:2009mm}. As a result, the pre-inflationary evolution is significantly changed, and the space of initial conditions capable of yielding sufficient slow-roll inflation is reshaped \cite{Ashtekar:2009mm,Ashtekar:2011rm}. Prior studies have shown that for minimally coupled scalar fields, LQC dynamics generally increase the likelihood of achieving successful inflation, although the degree of enhancement depends on the chosen potential and model details \cite{Ashtekar:2011rm,Bedic:2018gqu,Vereshchagin:2018dnm}.

In the effective description of LQC, quantum geometric effects arise primarily through two distinct mechanisms associated with the underlying discreteness of space. The first, known as holonomy  corrections, originates from representing the curvature (field strength) of the Ashtekar-Barbero connection by holonomies around loops of minimal nonzero area. These corrections are responsible for singularity resolution  via the ``Big Bounce'' and provide the dominant modification in the high-curvature (Planck) regime. Extensive studies have already analyzed the probability of inflation, focusing primarily on these corrections~\cite{Li:2023dwy, Barboza:2020jux, Graef:2018ulg, Bedic:2018gqu, Ashtekar:2011rm, Chen:2015yua}. The second mechanism, inverse volume corrections, arises from the quantization of operators involving inverse powers of the densitized triad (or equivalently the scale factor), which diverge classically but remain finite in LQC. In contrast to holonomy effects, inverse-volume modifications of the effective Hamiltonian and the equations of motion are not strictly confined to the deep Planck regime and may persist into the semi-classical phase~\cite{Bojowald:2007ky, Germani:2007rt}. While the effect of holonomy corrections on inflationary probabilities has been intensively investigated, a distinct gap remains regarding the specific impact of inverse volume corrections. Since these modifications may influence the phase-space flow during the onset of slow-roll inflation, in this work, we specifically focus our analysis on filling this gap in the literature.

A theoretically well-motivated extension of inflation is to include a non-minimal coupling between the inflaton and the Ricci scalar, typically written as $\xi\phi^{2}R$. In quantum field theory in curved spacetime, this operator is generically generated by renormalization and should therefore be included in the effective action, even if it is absent at the classical level \cite{Birrell:1982ix,Buchbinder:1992rb}. In an effective field theory formulation, it is likewise allowed by the symmetries and appears in the low-energy action with a coefficient that encodes ultraviolet sensitivity and the effects of integrating out heavy degrees of freedom \cite{Baumann:2014nda,Cheung:2007st}. Similar expectations apply in quantum-gravity motivated settings, where radiative and threshold effects typically induce such interactions in the low-energy effective action, as expected in ultraviolet completions such as string theory and higher-dimensional constructions \cite{Baumann:2014nda,Polchinski:1998rr}.

The non-minimal coupling has direct dynamical consequences. It makes the effective gravitational coupling field dependent in the Jordan frame and modifies the slow-roll conditions, thereby altering the relation between Lagrangian parameters and inflationary observables. Higgs inflation provides a representative example, since a sizable $\xi$ flattens the potential in the Einstein frame and allows the Higgs field to sustain slow-roll inflation within an economical setting connected to particle physics \cite{Bezrukov:2007ep,Bezrukov:2009db}. More broadly, scalar-tensor theories provide a systematic framework for analyzing these effects and confronting them with observations \cite{DeFelice:2010aj,Capozziello:2011et,Clifton:2011jh,Kobayashi:2019hrl,Lopez:2021agu,Leyva:2022zhz}.

A recent study confronted inverse-volume corrections in LQC with current data by considering loop quantum inflation driven by a canonical minimally coupled scalar field, with particular emphasis on low-scale spontaneously broken supersymmetric (SB SUSY) and exponential inflationary potentials \cite{Parvizi:2025sed}. In this work, we extend that analysis by incorporating inverse-volume corrections together with a non-minimal coupling $\xi\phi^{2}R$. Working at first order in the slow-roll expansion, we derive analytic expressions for the scalar spectral index $n_s$, the tensor-to-scalar ratio $r$, and the running $\alpha_s\equiv dn_s/d\ln k$, and we complement these results with numerical solutions of the corrected background evolution. We confront the resulting predictions with Planck 2018 \cite{Planck:2018jri} and ACT DR6 constraints \cite{AtacamaCosmologyTelescope:2025blo,AtacamaCosmologyTelescope:2025nti}. We also compute the probability of obtaining $N\ge N_\star$ e-folds after the bounce, defined through the Liouville measure on the effective phase space \cite{Ashtekar:2011rm,Bedic:2018gqu,Bojowald:2006hd,Han:2019mvj}, and assess how the non-minimal coupling modifies the phase-space volume of initial conditions leading to sufficient inflation.

We consider two classes of potentials. First, we study non-minimally coupled Higgs inflation with a quartic potential in the regime where the symmetry-breaking scale is negligible during inflation \cite{Bezrukov:2007ep}. Recent ACT DR6, SPT-3G, and DESI 2024 BAO results favor slightly larger $n_s$ than Planck 2018 \cite{Planck:2018jri} alone, while maintaining stringent upper bounds on the tensor-to-scalar ratio \cite{AtacamaCosmologyTelescope:2025blo,AtacamaCosmologyTelescope:2025nti,SPT-3G:2025bzu,DESI:2024uvr,DESI:2024mwx}. Depending on the dataset combination, this can place tension on plateau-type attractor models, including Starobinsky inflation and the standard Higgs-inflation limit, as reviewed in Ref.~\cite{Kallosh:2025ijd}. In this context, quantum-corrected $\phi^{4}$-type modifications of the effective potential, for instance from renormalization-group improved Higgs self-interactions, can bring Higgs inflation into improved agreement with the ACT-DESI preferred region \cite{Gialamas:2025kef,Yuennan:2025kde,Maity:2025czp}. We examine whether inverse-volume corrections in LQC can provide an additional shift of the predictions toward the region favored by current data.

Second, we analyze string-inspired fractional monomial potentials of the form $V(\phi)\propto\phi^{p}$, motivated by axion-monodromy and related ultraviolet constructions~\cite{Silverstein:2008sg,McAllister:2008hb,Martin:2013tda}. The preference for larger $n_s$ renews interest in these models, particularly for $p<1$, which can lie within the ACT-DESI preferred region \cite{Kallosh:2025rni,Maity:2025czp}. Together, these two cases allow us to assess how potential shape and non-minimal coupling interplay with inverse-volume corrections in determining both the inflationary predictions and the probability of sufficiently long inflation.

The paper is organized as follows. Section~\ref{sec:Inflation in a generalized non-minimally coupled gravity} reviews the background dynamics of inflation in generalized non-minimally coupled scalar–tensor theories. Section~\ref{sec:Effective dynamics in LQC with non-minimal coupling} discusses the construction of the Liouville measure and its quantum-corrected form in LQC. Section~\ref{sec:Higgs inflation with non-minimal coupling} analyzes Higgs inflation with non-minimal coupling, while Section~\ref{sec:String-inspired potentials with non-minimal coupling} extends the discussion to string-inspired fractional monomial potentials. Section~\ref{sec:Discussion_Conclusion} presents our conclusions. Throughout, we work in natural units $\hbar=c=1$, and $M_{Pl}$ denotes the reduced Planck mass.

\section{Inflation in a generalized non-minimally coupled gravity}
\label{sec:Inflation in a generalized non-minimally coupled gravity}

In this section, we briefly review the background dynamics and the linear theory of cosmological perturbations for inflation in a generalized scalar-tensor theory with a non-minimal coupling.

\subsection{Cosmological background evolution}
\label{subsec:Cosmological background evolution}

We start from the Jordan-frame action
\begin{equation}
    S=\int d^4x \sqrt{-g}\left[\frac{U(\phi)}{2\kappa}R - \frac{1}{2}\partial_\mu\phi\partial^\mu\phi-V(\phi)\right],
    \label{eq:classical_action}
\end{equation}
where $\kappa \equiv 1/M_{Pl}^2$.

This action describes a canonical scalar field $\phi$ with potential $V(\phi)$, non-minimally coupled to gravity through the coupling function $U(\phi)$  \cite{DeFelice:2010aj,DeFelice:2011jm,DeFelice:2011uc,Lopez:2021agu}.

For the background, we assume a spatially flat  Friedmann–Lemaître–Robertson–Walker (FLRW) spacetime 
\be
ds^2=-dt^2+a(t)^2\delta_{i j}dx^idx^j,
\label{eq:frlw_metric}
\ee where $a(t)$ is the scale factor. 
Substituting \eqref{eq:frlw_metric} into \eqref{eq:classical_action} and discarding total derivatives, one obtains the minisuperspace Lagrangian density \cite{Faraoni:2010pgm}
\begin{equation}
    \label{eq:frlw_lagrangian}
    \mathcal{L}=a^3 \left[\frac{1}{2}\dot{\phi}^2-V-\frac{3}{\kappa}\left(U H^2+\dot{U}H\right)\right],
\end{equation}
where $H\equiv \dot{a}/a$ is the Hubble rate, and $\dot{U}\equiv d U/d t$.

From \eqref{eq:frlw_lagrangian} we define the Hamiltonian density in the standard way,
\begin{equation}
\mathcal{H}\equiv \dot\phi\,\frac{\partial\mathcal{L}}{\partial\dot\phi}
+\dot a\,\frac{\partial\mathcal{L}}{\partial\dot a}
-\mathcal{L},
\end{equation}
which yields
\begin{equation}
    \label{eq:frlw_hamiltonian}
    \mathcal{H}=a^3 \left[\frac{1}{2}\dot{\phi}^2+V-\frac{3}{\kappa}\left(U H^2+\dot{U}H\right)\right].
\end{equation}
The Hamiltonian constraint $\mathcal{H}=0$ in \eqref{eq:frlw_hamiltonian} can be written as the first Friedmann equation,
\be
\label{eq:flrw_first_friedmann}
H^2=\frac{\kappa}{3U(\phi)}\left(\frac{1}{2}\dot\phi^{\,2}+V(\phi)\right)-H\frac{\dot U(\phi)}{U(\phi)}.
\ee
The second Friedmann equation can also be obtained directly from the mini-superspace Lagrangian by varying it with respect to the scale factor 
\begin{equation} \label{eq:flrw_friedmann}
   \frac{1}{2} \dot{\phi}^2 -V + \frac{U}{\kappa}\left( \frac{\ddot{a}}{a}+H^2 \right) + \frac{U^\prime}{\kappa}\left(\ddot{\phi}+2H\phi\right)=0.
\end{equation}

Similarly, the modified Klein-Gordon equation is obtained by varying with respect to $\phi$
\begin{equation}
    \label{eq:flrw_klein_gordon}
   \ddot{\phi} + 3 H \dot{\phi}+V^\prime - \frac{3U^\prime}{\kappa} \left( \frac{\ddot{a}}{a}+H^2 \right)=0,
\end{equation}
where a prime denotes the derivative with respect to $\phi$. The set of field equations \eqref{eq:flrw_first_friedmann}, \eqref{eq:flrw_friedmann}, and \eqref{eq:flrw_klein_gordon} constitutes the full system of cosmological equations for the non-minimally coupled scalar field model. However, only two of these field equations are independent.

To study the inflationary dynamics, we introduce the slow-roll parameters \cite{DeFelice:2011uc}
\bea
&& \epsilon=-\frac{\dot{H}}{H^2},\:\:\: \delta_{\phi}=\frac{\ddot{\phi}}{H\dot{\phi}}\:\:\: \delta_{X}=\frac{\kappa X}{H^2 U},\nonumber\\
&& \delta_{U}=\frac{\dot{U}}{H U},\:\:\: \delta_{\dot{U}}=\frac{\ddot{U}}{H \dot{U}},
\label{Slow_Para}
\eea
where $X=-\partial_{\mu}{\phi}\partial^{\mu}{\phi}/2$. During inflation, these parameters are assumed to be small, which defines the slow-roll approximation \cite{Baumann:2014nda,Lopez:2021agu,Leyva:2021fuo}.

Using these dimensionless parameters together with the background equations \eqref{eq:flrw_first_friedmann} and \eqref{eq:flrw_klein_gordon}, one gets
\bea
\epsilon=\delta_{X}-\frac{1}{2}\delta_{U}+\frac{1}{2}\delta_{\dot{U}}\delta_{U}.
\eea

To first order in slow-roll expansion, this reduces to
\be
\epsilon=\delta_{X}-\frac{1}{2}\delta_{U}+\mathcal{O}(\epsilon^2),
\label{epsilon}
\ee with $\delta_{X}, \delta_{U}\ll 1$.

Within the slow-roll approximation, the Friedmann and Klein-Gordon (KG) equations reduce to
\bea
&& H^2\simeq\frac{\kappa V}{3 U},\\
&& \frac{\dot{\phi}}{H}\simeq \frac{2 U^\prime}{\kappa}-\frac{V^\prime}{3 H^2}.
\label{phi_slow}
\eea

Therefore, to quantify the amount of inflation, we introduce the number of $e$-folds $N$ defined as
\bea
&& N\equiv \int_{t_{i}}^{t_f}{ H dt}=\int_{\phi_{i}}^{\phi_{f}}\frac{\kappa}{2 U^\prime-U\frac{V^\prime}{V}} d\phi.
\label{phi_N}
\eea 

In this latter equation, the field value at the end of inflation is calculated from the condition $\epsilon(\phi_{f})=1$ in \eqref{epsilon}, and $\phi_i$ denotes the field value at horizon crossing ($\phi_i\equiv \phi_*$).

 \subsection{Primordial Fluctuations}
\subsubsection{Scalar perturbations}
In order to study the evolution of primordial perturbations, we work in the unitary gauge, defined by $\delta{\phi}=0$. In this gauge, the scalar-perturbed FLRW metric can be written as \cite{DeFelice:2011uc}
\begin{eqnarray}
ds^{2} =-\left[ \left( 1+\alpha \right) ^{2}-a(t)^{-2}e^{-2\mathcal{R}}\left(
\partial \psi \right) ^{2}\right] dt^{2} 
+2\partial _{i}{\psi }dtdx^{i}+a(t)^{2}e^{2\mathcal{R}}\delta
_{ij}dx^{i}dx^{j}, 
\label{Pertubed_metric}
\end{eqnarray} 
where $\alpha$, $\psi$ are nondynamical modes, and $\mathcal{R}$ is the curvature perturbation. Here $\left(
\partial \psi \right) ^{2}\equiv \delta^{i j}\partial_{i}{\psi}\partial_{j}{\psi}$.

Substituting the perturbed metric \eqref{Pertubed_metric} into the action \eqref{eq:classical_action}, and expanding to second order, we integrate by parts and eliminate the nondynamical modes, obtaining
\be
S=\int{dt d^{3}x a^3 Q_{s}\left[\dot{\mathcal{R}}^2-\frac{c_{s}^2}{a^2}\left(\partial \mathcal{R}\right)^2\right]},
\label{SOA}
\ee 
where 
\be
Q_{s}=\frac{w_{1}\left(4 w_{1} w_{3}+9 w_{2}^2\right)}{3 w_{2}^2},\:\:\: c_{s}^2=\frac{3 \left(2 w_{1}^2 w_{2} H-w_{2}^2 w_{1}+4 w_{1}\dot{w}_{1} w_{2}-2 w_{1}^2 \dot{w}_{2}\right)}{w_{1}\left(4 w_{1} w_{3}+9w_{2}^2\right)}, 
\ee with
\begin{eqnarray}
w_{1} & = & \frac{U}{\kappa}\,,\\
w_{2} & = & \frac{2 HU}{\kappa}+\frac{\dot{\phi}}{\kappa}U^\prime,\\
w_{3} & = & -\frac{9H^{2}U}{\kappa}+3X
-\frac{9H\dot{\phi}}{\kappa} U^\prime.
\end{eqnarray}
To ensure the stability of scalar perturbations, we impose $Q_s>0$ (absence of ghosts) and $c_s^2>0$ (absence of gradient instabilities). Expanding $Q_s$ and $c_s^2$ to leading order in the slow-roll parameters, we obtain
\be
Q_s \simeq \frac{U\,\delta_X}{\kappa},
\qquad
c_s^2 \simeq 1.
\ee

After quantizing the curvature perturbation on a quasi-de Sitter background and solving the Mukhanov-Sasaki equation, the scalar power spectrum evaluated at sound-horizon crossing ($c_s k=aH$) is \cite{DeFelice:2011jm}
\bea
&& \mathcal{P}_{s}
=\left.\frac{H^2}{8 \pi^2 Q_{s} c_{s}^3}\right|_{c_{s}k=aH}
=\left.\frac{H^2}{8 \pi^2 M_{Pl}^2 U\, c_{s}\epsilon_{s}}\right|_{c_{s}k=aH},\nonumber\\
&& \simeq \left.\frac{\kappa H^2}{8 \pi^2 U\delta_{X}}\right|_{c_{s}k=aH}, 
\eea where we have also defined 
\be \label{eq:new_roll}
\epsilon_{s}\equiv \frac{\kappa Q_{s}c_{s}^2}{U}\simeq \delta_{X}.
\ee
The corresponding spectral index reads
\be
n_{s}-1\equiv \left.\frac{d\ln \mathcal{P}_{s}}{d \ln k}\right|_{c_{s}k=aH}
\simeq -2\epsilon-\delta_{U}-\eta_{s},
\ee
where
\be
\eta_{s}\equiv \frac{\dot{\epsilon}_{s}}{H \epsilon_{s}}.
\ee Here $\eta_s$ quantifies the fractional variation of $\epsilon_s$ per Hubble time. The spectral tilt is evaluated at sound-horizon crossing, keeping only first-order slow-roll contributions.

Also, a useful parameter is the running of the scalar spectral index, which is defined as
\be
\alpha_{s}\equiv\left.\frac{dn_{s}}{d\ln{k}}\right|_{c_{s}k=aH}
\label{def_running},
\ee whose value is also constrained by the latest observational data.

\subsubsection{Tensor perturbations}
For tensor perturbations, we consider the transverse-traceless (TT) deformation of the spatial metric \cite{DeFelice:2011uc},
\be
ds^{2}=-dt^{2}+a^{2}(t)\left(\delta_{ij}+h_{ij}\right)dx^{i}dx^{j},
\ee
where $h_{ij}$ satisfies
\be
\partial^{i}h_{ij}=0,\qquad h^{i}{}_{i}=0.
\ee
Tensor perturbations can be decomposed into the two polarization states as \cite{Baumann:2014nda}
\be
h_{ij}=h_{+}e^{+}_{ij}+h_{\times}e^{\times}_{ij},
\ee
which leads to the quadratic action
\be
S_{t}^{(2)}=\sum_{p=+,\times}\int dt\, d^{3}x\, a^{3}Q_{t}\left[\dot{h}_{p}^{2}-\frac{c_{t}^{2}}{a^{2}}\left(\partial h_{p}\right)^{2}\right],
\ee
with
\be
Q_{t}=\frac{w_{1}}{4},\qquad c_{t}^{2}=1.
\ee
Hence, tensor modes propagate at the speed of light and are free of gradient instabilities, while the no-ghost condition requires $Q_{t}>0$.
The tensor power spectrum, evaluated at horizon crossing, is \cite{DeFelice:2011uc}
\be
\mathcal{P}_{t}=\left.\frac{H^{2}}{2\pi^{2}Q_{t}c_{t}^{3}}\right|_{c_{s}k=aH}.
\ee
The corresponding tensor tilt is
\be
n_{t}\equiv \left.\frac{d\ln \mathcal{P}_{t}}{d\ln k}\right|_{c_{s} k=aH}=-2\epsilon-\delta_{U}.
\ee
Finally, the tensor-to-scalar ratio reads
\be
r=\frac{\mathcal{P}_{t}}{\mathcal{P}_{s}}=16c_{s}\epsilon_{s}
\simeq -8 n_{t},
\ee
showing that $r$ and $n_t$ are not independent.

\section{Effective dynamics in LQC with non-minimal coupling}
\label{sec:Effective dynamics in LQC with non-minimal coupling}

\subsection{Effective equations and bounce}
In loop quantum cosmology, it is convenient to work with the canonical pair adapted to isotropic
geometries~\cite{Agullo:2013dla,Bojowald:2005epg,Bojowald:2004ax,Gambini:2011zz,Thiemann:2007pyv}
\begin{equation}
    |p| := a^2, \qquad c := \gamma \dot{a},
\end{equation}
where $\gamma$ is the Barbero-Immirzi parameter. For later convenience, we also introduce 
\begin{equation}
    \pi := a^3 \dot{\phi} = p^{3/2} \dot{\phi},
\end{equation}
which should not be confused with the canonical momentum of $\phi$,
$\pi_\phi=\partial\mathcal{L}/\partial\dot\phi=\pi -(3/\kappa) a^3 \ H \ U^\prime$, in the presence of the non-minimal coupling.

In terms of $(p,c,\phi,\pi)$, the mini-superspace Lagrangian can be written as
\begin{equation}
    \mathcal{L}=-\frac{3}{\kappa\gamma^2}\sqrt{p}c^2 U-\frac{3}{\kappa\gamma}pc U^\prime\dot{\phi}+\frac{\pi^2}{2p^{3/2}}-p^{3/2}V,
\end{equation}
leading to the Hamiltonian
\begin{equation}\label{eq:pre_hamiltonian}
\mathcal{H}=-\frac{3}{\kappa\gamma^2}\sqrt{p}c^2 U-\frac{3}{\kappa\gamma}pc U^\prime\dot{\phi}+\frac{\pi^2}{2p^{3/2}}+p^{3/2}V.
\end{equation}
Quantum-geometry effects are incorporated at the effective level by (i) implementing holonomy
corrections, $c\rightarrow \sin(\mu_0 c)/\mu_0$ ~\cite{Artymowski:2012is, Bojowald:2006hd}, where $\mu_0$ represents the ``length'' of the edge along which the holonomy is computed, and (ii) including inverse-volume corrections encoded in
the function $D_l(q)$~\cite{Germani:2007rt,Bojowald:2001xe,Bojowald:2004xq,Bojowald:2002ny}. The resulting effective Hamiltonian reads
\begin{equation}\label{eq:classical_hamiltonian}
    \mathcal{H}_{LQ}=-\frac{3}{\kappa\gamma^2\mu_0^2}\sqrt{p_{LQ}}\sin^2{\left(\mu_0 c_{LQ}\right)}U-\frac{3}{\kappa\gamma^2\mu_0}p_{LQ}\sin{\left(\mu_0 c_{LQ}\right)}U^\prime\dot{\phi}_{LQ}+D_l^{-(n+1)}\frac{\pi_{LQ}^2}{2p_{LQ}^{3/2}}+D_l^{m}p_{LQ}^{3/2}V,
\end{equation}
Here, $m$ and $n$ are operator-ordering parameters whose precise definition  and physical role are given below in Eq.~\eqref{eq:ordering}.

The inverse-volume factor arises from the expectation values of the volume and inverse-volume
operators,
\begin{equation}
    \langle\widehat{a^3}\rangle=a_{LQ}^3, \qquad \langle\widehat{a^{-3}}\rangle=D_l(q) \ a_{LQ}^{-3},
\end{equation}
with $q\equiv a_{LQ}^{2}/a_{\rm Pl}^{2}$ and $a_{\rm Pl}$ denoting the scale factor at the quantum-to-classical
transition. The explicit form of $D_l(q)$ is \cite{Germani:2007rt}
{\small
\be
D_l(q) = \left\{ \frac{3q^{1-l}}{2l} \left[ \frac{(q+1)^{l+2} - |q-1|^{l+2}}{l+2} - \frac{q}{1+l} \left( (q+1)^{l+1} - \text{sgn}(q-1)|q-1|^{l+1} \right) \right] \right\}^{\frac{3}{2-2l}}.
\ee}
The parameter $l\in(0,1)$ is an ambiguity parameter associated with the quantization of inverse powers of the triad. It enters the inverse-volume correction function $D_l(q)$ and controls how rapidly this function approaches its classical value in the semiclassical regime. The value $l=3/4$ is not selected by a fundamental principle of LQC. Rather, it has become a standard fiducial choice in the inverse-volume-correction literature \cite{Bojowald:2002ny,Germani:2007rt} because it yields a well-behaved correction function with a smooth quantum-to-classical transition and allows direct comparison with previous studies of loop quantum inflation. In this sense, $l=3/4$ should be understood as a representative benchmark choice, not as a unique prediction of the theory. A full exploration of the dependence on $l\in(0,1)$ would correspond to a separate analysis of quantization ambiguities.

\begin{figure}[H]
    \centering
    \begin{minipage}{0.5\linewidth}
         \centering
        \includegraphics[width=0.9\textwidth]{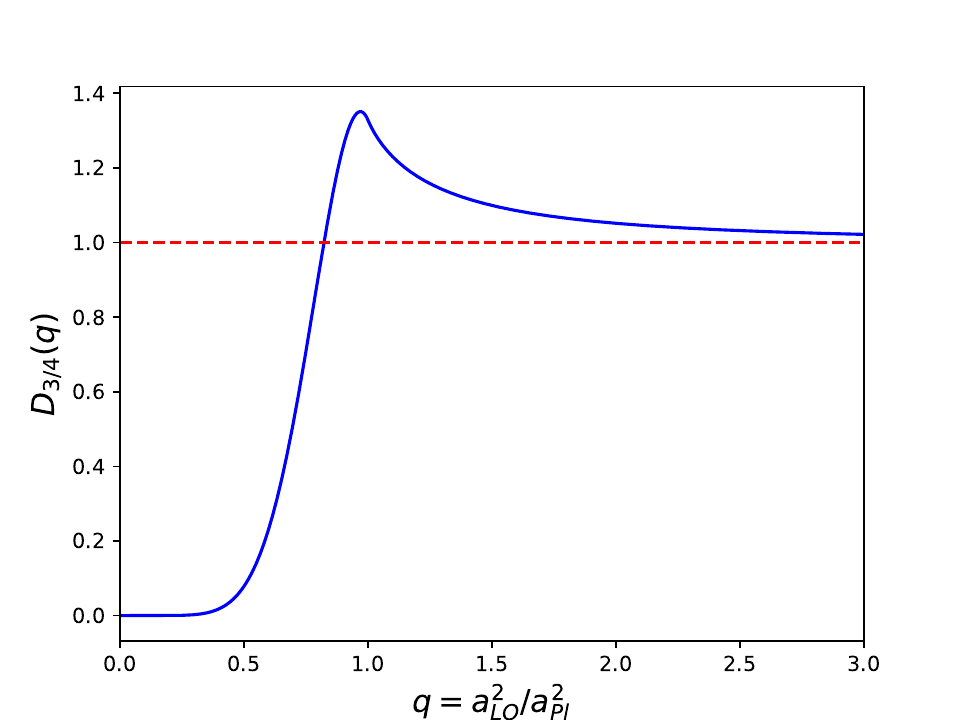}
    \end{minipage}\hfill
    \begin{minipage}{0.5\linewidth}
        \centering
        \includegraphics[width=0.9\textwidth]{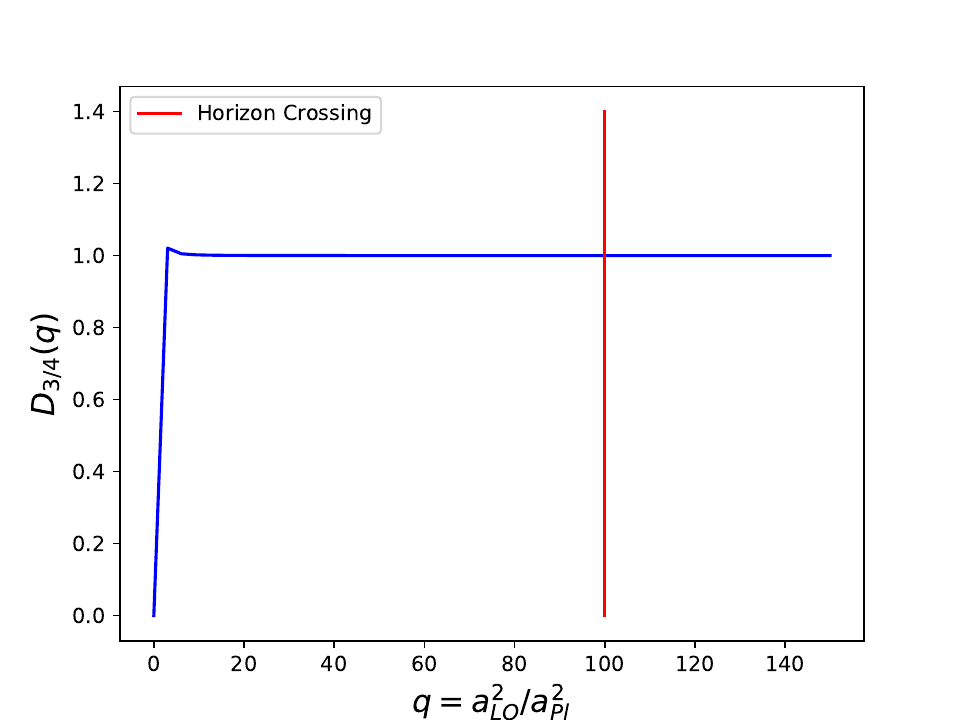}
    \end{minipage}\\
    \caption{Inverse-volume correction factor $D_l(q)$ as a function of the  dimensionless variable $q \equiv a_{LQ}^2/a_{\rm Pl}^2$, shown for the standard eigenvalue parameter $l=3/4$. Left panel: behavior in the narrow domain $q \in [0,3]$, illustrating the strong deviation of $D_l(q)$ from its classical value near the quantum bounce ($q \sim 1$) and the rapid quantum-to-classical transition. Right panel: asymptotic behavior over the  broader range $q \in [0,150]$. The red vertical line marks the location of horizon crossing at $q_i=100$, confirming that the primordial perturbations are generated deep in the quasi-classical regime ($q \gg 1$), where $D_l \to 1$ and the inverse-volume corrections are small and perturbative.}
    \label{fig:Dl_plot}
\end{figure}

Upon quantization, operator-ordering ambiguities allow one to split powers of $a$ in different ways.
Parametrizing this freedom by $(m,n)$, one finds schematically \cite{Germani:2007rt}
\be
\begin{split} \label{eq:ordering}
    a_{LQ}^3=a_{LQ}^{3(m+1)}a_{LQ}^{-3m}&\to \langle\widehat{a^3}\rangle^{(m+1)} \ \langle\widehat{a^{-3}}\rangle^m = D_l^m a_{LQ}^3,\\
    a_{LQ}^{-3}=a_{LQ}^{-3(n+1)}a_{LQ}^{3n}&\to \langle\widehat{a^{-3}}\rangle^{(n+1)} \ \langle\widehat{a^3}\rangle^{n} = D_l^{-(n+1)} a_{LQ}^3.
\end{split}
\ee
The parameters $m$ and $n$ are not fixed by a unique ordering prescription, but their physical impact is entirely governed by the profile of the correction factor $D_l(q)$. In the deep quantum regime near the bounce, $D_l(q)$ deviates strongly from its classical limit. This behavior is illustrated in the left panel of FIG.~\ref{fig:Dl_plot} for the standard parameter choice $l=3/4$ over the narrow domain $q \in [0,3]$, clearly highlighting the quantum-to-classical transition that occurs around $q=1$. Conversely, in the classical macroscopic regime ($q\gg 1$), the discrete quantum geometry is coarse-grained, and one recovers $D_l\to 1$, naturally switching off the corrections. The right panel demonstrates this asymptotic behavior over a broader range up to $q=150$. Crucially, in our cosmological setup, the horizon crossing of observable scales occurs around $q=100$. This confirms that the primordial perturbations are generated deep within the quasi-classical regime, where the inverse-volume effects provide small, perturbative modifications rather than dominating the background dynamics.

At the quantum level, physical states satisfy the Hamiltonian constraint
\begin{equation}
    \widehat{\mathcal{H}}_{LQ} \ket{\Phi}=0,
\end{equation}
with
\begin{equation}\label{eq:hamiltonian}
    \begin{split}
        \widehat{\mathcal{H}_{LQ}}=&-\frac{3}{\kappa\gamma^2\mu_0^2} \ \left(\widehat{p^{3/2}}\right)^{1/3} \ \left(\widehat{\sin{\left(\mu_0 c\right)}}\right)^2 \ \widehat{U}-\frac{3}{\kappa\gamma^2\mu_0}\left(\widehat{p^{3/2}}\right)^{1/3} \ \widehat{\sin{\left(\mu_0 c\right)}} \ \widehat{U^\prime} \ \widehat{\pi}\\
        &+\left(\widehat{p^{-3/2}}\right)^{(n+1)} \ \left(\widehat{p^{3/2}}\right)^{n}  \ \frac{\widehat{\pi}^2}{2} + \left(\widehat{p^{3/2}}\right)^{(m+1)} \ \left(\widehat{p^{-3/2}}\right)^{m} \ \widehat{V}.
    \end{split}
\end{equation}
Here, the parameters $m$ and $n$ encode the usual quantization ambiguity associated with how
powers of the scale factor are distributed among operators in the prefactors of the kinetic and
potential terms. In addition, the gravitational sector entails an operator-ordering ambiguity because the triad and holonomy operators do not commute. In particular,

\begin{equation}
    [\widehat{p}, \widehat{e^{i\mu_0 c}}] = \frac{8\pi \gamma \ell_P^2 \mu_0}{3} \widehat{e^{i\mu_0 c}}, \qquad [\widehat{\phi},\widehat{\pi}_\phi]=i,
\end{equation}
so different orderings of $\widehat{p}$ and holonomy operators (or $\widehat{p}$ and $\widehat{\pi}$) lead to inequivalent realizations of the quantum constraint.
Although this second source of ambiguity is typically negligible at the characteristic energy scales of slow-roll inflation, it plays a crucial role in the deep quantum regime, where the detailed operator ordering influences the construction of the physical state space and the definition of the physical vacuum state.

In the classical limit, one recovers the standard Hamiltonian constraint $\langle\widehat{\mathcal{H}}\rangle=0$ by taking
$D_l\to 1$ and $\sin(\mu_0 c)\to \mu_0 c$, together with the replacement of operators by their
classical counterparts (and commutators by Poisson brackets).

\subsection{Slow-roll regime and number of e-folds}

We restrict ourselves to the quasi-classical regime $\mu_0|c|\ll 1$, so that $\sin(\mu_0 c)\simeq\mu_0 c$ and holonomy corrections can be neglected. It is worth elaborating on the relative importance of the two types of quantum geometric corrections across different dynamical regimes. Holonomy corrections, which arise from replacing the connection component $c$ by $\sin(\mu_0 c)/\mu_0$ in the effective Hamiltonian, are the dominant modification in the deep Planck regime, where the energy density approaches the critical density $\rho_c \sim M_{\rm Pl}^4$. They are directly responsible for the resolution of the big-bang singularity via the quantum bounce and drive the superinflationary phase immediately after the bounce \cite{Ashtekar:2009mm, Copeland:2007qt}. However, as the universe expands away from the bounce and the energy density drops well below $\rho_c$, the holonomy corrections decay rapidly as $(\rho/\rho_c)$ and become completely negligible in the slow-roll inflationary regime \cite{Ashtekar:2011rm}. Inverse-volume corrections, by contrast, arise from the quantization of operators involving inverse powers of the densitized triad, and their characteristic scale is set by the quantum-to-classical transition at $q \sim 1$ rather than by the Planck density. As a result, they are not strictly confined to the deep Planck regime and may persist into the semi-classical phase \cite{Bojowald:2007ky, Germani:2007rt}, providing perturbatively small but non-negligible modifications to the slow-roll dynamics and the primordial power spectra, as encoded in the correction factor $D_l(q)$. In our cosmological setup, horizon crossing occurs at $q_i = 100$, deep in the quasi-classical regime where $D_l(q_i) \gtrsim 1$ but is close to unity. In this regime, holonomy corrections are entirely negligible, while inverse-volume corrections remain the leading quantum geometric effect. This justifies our focus on inverse-volume corrections alone throughout this work, and in this limit, the only LQC corrections we retain are the inverse-volume effects encoded in $D_l(q)$, which may still be relevant near the quantum-to-classical transition.

In this limit, the effective Hamiltonian constraint reduces to
\begin{equation}\label{eq:hamiltonian_constraint}
    -\frac{3}{\kappa}H_{LQ}\left[H_{LQ}U+\dot{\phi}_{LQ}U^\prime\right]+D_l^{-(n+1)}\frac{\dot{\phi}_{LQ}^2}{2}+D_l^m V=0,
\end{equation}
which plays the role of the LQC-modified Friedmann equation.

It is also convenient to work with the corresponding effective Lagrangian,
\begin{equation}
    \label{eq:lagrangian}
    \mathcal{L}_{LQ}=a_{LQ}^3\left[-\frac{3}{\kappa}H_{LQ}\left(H_{LQ}U+\dot{\phi}_{LQ}U^\prime\right)+D_l^{-(n+1)}\frac{\dot{\phi}_{LQ}^2}{2}-D_l^m V\right],
\end{equation}
from which we obtain the LQG-corrected Klein-Gordon equation
\begin{equation}\label{eq:klein_gordon}
    \begin{split}
        - \frac{3U^\prime}{\kappa} \left[ \left(\frac{\ddot{a}}{a}\right)_{LQ}+H_{LQ}^2 \right] + D_l^{-(n+1)}\left[\ddot{\phi}_{LQ} + 3 H_{LQ} \ \dot{\phi}_{LQ}-(n+1)\left(\frac{\dot{D_l}}{D_l}\right)\dot{\phi}_{LQ}\right]+D_l^m V^\prime =0,
    \end{split}
\end{equation}
together with the second (Raychaudhuri-type) Friedmann equation
\begin{equation}\label{eq:2nd_firedmann}
    \frac{1}{\kappa}\left[2U\left(\frac{\ddot{a}}{a}\right)_{LQ}+H_{LQ}\left(2U^\prime\dot{\phi}_{LQ}+H_{LQ}U\right)+\dot{\phi}_{LQ}^2 U^{\prime\prime}+\ddot{\phi}_{LQ}U^\prime\right]+D_l^{-(n+1)}\frac{\dot{\phi}_{LQ}^2}{2}-D_l^m V = 0.
\end{equation}

Implementing the LQC-modified slow-roll conditions in \eqref{Slow_Para} and defining
\begin{equation}
    \label{slow_lqg}
    \begin{split}
        \left(\delta_D\right)_{LQ}&\equiv \frac{\dot{D}_l}{H_{LQ}D_l}=\delta_D\\
        &= \frac{3 \left[(q+1)^l \left(l^2-3 l q+3 q^2-1\right)-(q-1)^l \left(l^2+3 l q+3 q^2-1\right)\right]}{(l-1) \left[(l+q+1) (q-1)^{l+1}+(l-q+1) (q+1)^{l+1}\right]},
    \end{split}
\end{equation}

\begin{figure}[H]
    \centering
    \includegraphics[width=0.5\textwidth]{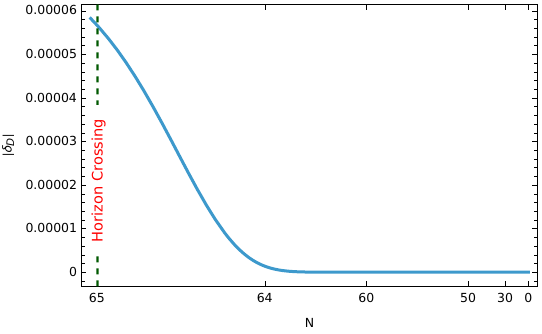}
    \caption{Evolution of the dimensionless parameter $\delta_D \equiv \dot{D}_l/(H_{LQ} D_l)$ as a function of the number of e-folds $N$, evaluated for $l=3/4$ and initial value $q_i=100$. The parameter $\delta_D$ quantifies the fractional rate of change of the inverse-volume correction factor $D_l(q)$ per Hubble time. It is dynamically significant only during the immediate post-bounce quantum-to-classical transition near $q \sim 1$ (large $N$), and asymptotically vanishes ($\delta_D \to 0$) as the universe enters the inflationary regime ($q \gg 1$). The dark green dashed vertical line marks the location of horizon crossing, showing that $|\delta_D|$ is already negligibly small throughout the observable e-folding window, justifying its treatment as a first-order slow-roll parameter.}
    \label{fig:delta_D}
\end{figure}

we get the $0^{\text{th}}$ order Friedmann equation\footnote{Note that the unscripted quantities on the right-hand side correspond to their classical functional forms, but are evaluated along the effective LQC trajectory $\phi_{LQ}$ i.e, $H_{LQ}^2 (\phi_{LQ}) =  D_l^m H^2(\phi_{LQ})$.}
\begin{equation}\label{eq:friedmann_zero}
    H_{LQ}^2=\kappa D_l^m \frac{V}{3U}=D_l^m H^2,
\end{equation}
and the $0^{\text{th}}$ order Klein-Gordon equation
\begin{equation}\label{eq:klein_gordon_zero}
    \frac{\dot{\phi}_{LQ}}{H_{LQ}}=\frac{1}{3}D_l^{-(n+1)}\left[\frac{6U^\prime}{\kappa}-\frac{D_l^m V^\prime}{H_{LQ}^2}\right]=D_l^{-(n+1)}\frac{\dot{\phi}}{H}.
\end{equation}

As illustrated in FIG.~\ref{fig:delta_D}, the parameter $\delta_D$ captures the fractional rate of change of the inverse-volume correction over a Hubble time. In the immediate post-bounce phase, during the quantum-to-classical transition near $q \sim 1$, $D_l(q)$ varies rapidly (as shown in the left panel of FIG.~\ref{fig:Dl_plot}), making $\delta_D$ a dynamically significant quantity. However, as the universe expands and successfully enters the inflationary regime ($q \gg 1$), the effective geometry smooths out, and $D_l$ approaches unity (right panel of FIG.~\ref{fig:Dl_plot}). Consequently, its time derivative naturally diminishes, and $\delta_D$ asymptotically vanishes ($\delta_D \to 0$), as confirmed by FIG.~\ref{fig:delta_D}, where $|\delta_D|$ is already negligibly small well before horizon crossing. Because both $\delta_D$ and its time variation over a Hubble time remain sufficiently small throughout the observable $e$-folding window, $\delta_D$ behaves mathematically just like a standard kinematic slow-roll parameter, allowing it to be consistently incorporated into the first-order perturbative expansion.

These expressions, Eq. \eqref{eq:friedmann_zero} and Eq. \eqref{eq:klein_gordon_zero}  allow us to relate the inflaton evolution to the e-folding number $N_{LQ}$ via
\begin{equation}\label{eq:dPhi_dN}
    \begin{split}
        \frac{d\phi_{LQ}}{dN_{LQ}} &=\frac{1}{3}D_l^{-(n+1)}\left[\frac{6U^\prime}{\kappa}-\frac{D_l^m V^\prime}{H_{LQ}^2}\right], \\
        N_{LQ} &=\int_{\left(\phi_{i}\right)_{LQ}}^{\left(\phi_{f}\right)_{LQ}}\frac{3D_l^{(n+1)}}{\left[\frac{6U^\prime}{\kappa}-\frac{D_l^m V^\prime}{H_{LQ}^2}\right]} \ d\phi_{LQ}.
    \end{split}
\end{equation}
This relation will be used in the next section to evaluate the inflationary
observables at horizon crossing, by expressing all background quantities in terms of the field value
(or equivalently $N_{LQ}$).

\subsection{Perturbations and observables}

Starting from the second order action \eqref{SOA} and retaining only inverse-volume corrections, we obtain 
\be
\label{Qs_cs}
\begin{split}
    \left(Q_{s}\right)_{LQ} &\simeq \left(\frac{U \delta_{X}}{\kappa}\right)_{LQ}=D_l^{-2(n+1)}\frac{U \delta_{X}}{\kappa}=D_l^{-2(n+1)}Q_{s},\\
    \left(c_{s}\right)_{LQ}^2 &\simeq D_l^{-(n+1)} \left( 1 - \frac{n+1}{2} \delta_D \right).
\end{split}
\ee
These modifications to the perturbation amplitude can be rigorously understood through the ``dressed metric'' approach to LQC \cite{Agullo:2013dla}. In this framework, rather than deforming the constraint algebra of the perturbations, the linear scalar and tensor modes are treated as quantum test fields propagating on a coarse-grained, quantum-corrected effective background. Because the quantum fluctuations of the geometry are traced over to produce a dressed effective metric that preserves the standard Lorentzian causal structure, the perturbations strictly propagate at the speed of light, yielding $\left(c_s\right)_{LQ}^2 = 1$. Furthermore, even within the alternative anomaly-free framework where inverse-volume effects explicitly deform the constraint algebra \cite{Bojowald:2007ky, Parvizi:2025sed}, the modified speed of sound is given by Eq. \eqref{Qs_cs}. Since the inflationary observables are evaluated at horizon exit deep in the slow-roll regime ($q\gg 1$), the inverse-volume contributions that enter the propagation sector approach their classical limit ($D_l\to 1$ and $\delta_D\to 0$), so any deviation of $(c_s)_{LQ}^2$ from unity is higher order. Therefore, at first order, it is consistent in both implementations to adopt an effective $\left(c_s\right)_{LQ}^2 \simeq 1$ when computing the primordial power spectra.

Accordingly, the slow-roll parameters entering the spectra rescale as

\bea
&& \left(\delta_{U}\right)_{LQ}=D_l^{-(n+1)}\delta_{U},\\
&& \left(\delta_{X}\right)_{LQ}=D_l^{m-2(n+1)}\delta_{X}.
\eea

and the quantity $\eta_s$ becomes

\bea
&& \left(\eta_{s}\right)_{LQ}= D_l^{-m/2}\eta_{s}-2(n+1)\delta_D.
\eea

The first slow-roll parameter is, therefore,

\be
\epsilon_{LQ} \simeq \left(\delta_{X}\right)_{LQ}-\frac{1}{2}\left(\delta_{U}\right)_{LQ} =D_l^{m-2(n+1)}\delta_{X}-\frac{1}{2}D_l^{-(n+1)}\delta_{U}.
\ee

As a result, the scalar power spectrum at horizon crossing $(\phi_{LQ},q)=\left(\left(\phi_i\right)_{LQ}, q_i\right)$ becomes
\be
\left(\mathcal{P}_{s}\right)_{LQ} = D_l^{m+2(n+1)} \mathcal{P}_{s}\Bigg|_{\left(\phi_i\right)_{LQ}, \ q_i}.
\ee
Therefore, the scalar spectral index reads

\be
\begin{split}
    \left(n_s\right)_{LQ}-1 &= -2 \epsilon_{LQ}-\left(\delta_{U}\right)_{LQ}- \left(\eta_{s}\right)_{LQ}\Bigg|_{\left(\phi_i\right)_{LQ}, \ q_i}, \\
    &= -2D_l^{-(n+1)}\delta_X - D_l^{-m/2}\eta_s+2(n+1)\delta_D\Bigg|_{\left(\phi_i\right)_{LQ}, \ q_i}.
\end{split}
\ee
Finally, the tensor-to-scalar ratio is corrected according to
\be
r_{LQ} = D_l^{-2(n+1)}r\Bigg|_{\left(\phi_i\right)_{LQ}, \ q_i}.
\ee

The inverse-volume corrections modify the inflationary observables through powers of $D_l(q_i)$ evaluated at horizon crossing: the scalar spectral index $(n_s)_{LQ}$ receives corrections through the rescaled slow-roll parameters $(\delta_X)_{LQ}$, $(\delta_U)_{LQ}$, and $(\eta_s)_{LQ}$, while the tensor-to-scalar ratio $r_{LQ}$ is suppressed by the factor $D_l^{-2(n+1)}$. Since $D_l(q_i) \lesssim 1$ at $q_i=100$, these corrections are perturbatively small but shift the predictions in a direction that can improve agreement with current CMB data.

Having established the effective post-bounce dynamics, we now assess the typicality of a sufficiently
long slow-roll phase by using the Liouville measure on the space of solutions to compute the fraction
of trajectories with $N\ge N_\star$.

\subsection{Probability of inflation from the induced Liouville measure}
\label{sec:Probability}
 The probability of inflation is defined using the Liouville measure and its induced measure on the constraint
surface after gauge fixing \cite{Schiffrin:2012zf,Gibbons:2006pa}. Details are given in
~\ref{app_1_sec:measure}. We parametrize the effective constraint surface by
$Q^i=\{\phi,\ q\equiv a^2/a_{\rm Pl}^2,\ H\}$.

For the effective Lagrangian in Eq. \eqref{eq:lagrangian}, the conjugate momenta are defined as
\begin{align} \label{eq:mometum}
    P_{Q^i} = \frac{\partial \langle \mathcal{L}_{LQ} \rangle}{\partial \dot{Q^i}}.
\end{align}
In particular, one finds
\begin{align} \label{eq:mometum_phi}
    \begin{split}
        P_{\phi_{LQ}} &= a_{\text{Pl}}^3 q^{3/2} \left(D_l^{-(n+1)}\dot{\phi}_{LQ}-\frac{3}{\kappa}H_{LQ}U^\prime\right),\\
        P_{H_{LQ}} &= 0.
    \end{split}
\end{align}

Following Refs.~\cite{Nelson:2007fd,Germani:2007rt,Ashtekar:2011rm}, we introduce the
vector potential 
\begin{equation}\label{eq:vector_potential}
    \mathbf{A}=(P_{\phi_{LQ}},P_{H_{LQ}},P_q),
\end{equation}
and the associated divergence-free field
\begin{align}\label{eq:divergenceless_field}
    \mathbf{B} = \mathbf{\nabla} \times \mathbf{A}.
\end{align}
This formulation is equivalent to pulling back the Liouville measure to the Hamiltonian constraint surface and
evaluating the induced measure on a transversal section. Restricting to the constant-$q$ hypersurface $q=q_s$,
the corresponding measure element can be written as the flux element $B_q\,\mathrm{d}H_{LQ}\,\mathrm{d}\phi_{LQ}$ \cite{Schiffrin:2012zf,Gibbons:2006pa}.

The normalization factor is then defined as
\begin{equation}\label{eq:normalisation_factor}
    \mathcal{N}=\int\int B_q \ \mathrm{d}H_{LQ} \ \mathrm{d}\phi_{LQ},
\end{equation}

where $B_q$ denotes the $q$-component of $\mathbf{B}$. The probability of inflation is given by the ratio \cite{Gibbons:2006pa, Gibbons:1986xk}

\begin{equation}
    \mathcal{P}=\frac{\mathcal{M}}{\mathcal{N}},
\end{equation}
with
\begin{equation}\label{eq:probability_factor}
    \mathcal{M}=\int\int B_q \ \mathrm{d}H_{LQ} \ \mathrm{d}\phi_{LQ} \Bigg|_{\text{inflation}}.
\end{equation}
For our effective dynamics, one obtains
\begin{equation}\label{eq:B_q}
    B_q= a_{\text{Pl}}^3 q^{3/2}  \left[\frac{D_l^{m} V'}{3 H_{LQ}^2}-\frac{U^\prime}{\kappa} \right].
\end{equation}
Equivalently, the probability of obtaining an inflationary phase within a specified range of initial data can be
written as \cite{Bedic:2018gqu}
\begin{equation}\label{eq:probability_final_exp}
    \mathcal{P}(N_{LQ},\phi_{LQ})=\frac{1}{Z}\left| \int_{\left(\phi_{i}\right)_{LQ}}^{\left(\phi_f\right)_{LQ}}\int_{\left(H_i\right)_{LQ}}^{\left(H_f\right)_{LQ}} \ B_q \ dH_{LQ} \ d\phi_{LQ}\right|,
\end{equation}
where $Z$ is a normalization constant  (of the same nature as $\mathcal{N}$).

A few remarks on the choice of measure and its physical interpretation are in order. The Liouville measure adopted here is the canonical measure induced on the constraint surface of the effective Hamiltonian by the symplectic structure of the phase space, as detailed in ~\ref{app_1_sec:measure}. Its key virtue is that it satisfies three fundamental requirements: it is positive, it is independent of the choice of parametrization or the particular hypersurface used to evaluate it, and it respects the symmetry of the solution space without introducing additional ad hoc structure \cite{Gibbons:1986xk, Gibbons:2006pa, Schiffrin:2012zf}. In particular, the gauge-fixed version of this measure, based on the divergence-free field $\mathbf{B} = \nabla \times \mathbf{A}$ constructed from the symplectic form (Eq.~\eqref{eq:divergenceless_field}), identifies nearly flat universes that are physically indistinguishable, thereby avoiding the divergence that afflicts the naive Liouville measure \cite{Gibbons:2006pa}. The physical interpretation of the resulting probability is that of a typicality measure: it quantifies the fraction of distinct classical histories — labeled by their initial conditions on the constraint surface — that give rise to a slow-roll inflationary phase of at least $N_\star$ e-folds. It does not assign a probability to quantum events or to the bounce itself, but rather characterizes the distribution of post-bounce trajectories inherited from the effective LQC dynamics. We note that alternative measure choices, such as the volume-weighted measure or measures based on different gauge-fixing conditions, can in principle yield different numerical values for the probability \cite{Schiffrin:2012zf}. However, the qualitative trend identified in this work — namely the attractor-like enhancement of the inflationary probability with increasing (or decreasing) $\xi$ and its eventual saturation — is a robust feature of the phase-space flow that is not expected to depend sensitively on the specific measure choice, since it reflects the global structure of the slow-roll basin of attraction rather than fine details of the weighting.

In the following section we specialize this probabilistic framework to Higgs inflation with a non-minimal
coupling. We derive the corresponding effective background dynamics and inflationary observables, and then use the
induced Liouville measure introduced above to quantify the fraction of trajectories that produce a sufficiently
long inflationary phase.

Only the inverse-volume corrected quantities will be evaluated for the analysis. Henceforth we will drop the subcript ``${LQ}$'' for convinience.

\section{Higgs inflation with non-minimal coupling}
\label{sec:Higgs inflation with non-minimal coupling}

We consider the standard non-minimal coupling function
\begin{equation}\label{eq:non_minimal_coupling}
    U(\phi)=1+\xi\phi^2,
\end{equation}
together with a Higgs-like quartic potential
\begin{equation}\label{eq:higgs_potential}
    V(\phi)=\lambda(\phi^2-v_0^2)^2.
\end{equation}
Identifying the inflaton field with the Higgs field as the inflaton is motivated by minimality, since it utilizes the only fundamental scalar field already observed in nature, requiring no new particle degrees of freedom beyond the Standard Model. In the absence of a non-minimal coupling ($\xi=0$), the quartic potential $V(\phi) \propto \phi^4$ is disfavored by current CMB constraints, as it generically predicts a tensor-to-scalar ratio $r$ that is too large. However, introducing a non-minimal coupling term $\xi \phi^2 R$ effectively flattens the potential in the Einstein frame, leading to values of $(n_s,r)$ compatible with observations. Furthermore, recent ACT DR6 results \cite{AtacamaCosmologyTelescope:2025blo,AtacamaCosmologyTelescope:2025nti,Ellis:2025zrf}, particularly when combined with Planck and DESI \cite{Kallosh:2025rni,Maity:2025czp,Kumar:2025apf}, mildly favor a slightly
larger scalar spectral index (with a representative value $n_s\simeq 0.974$). This upward shift can introduce
tension for standard plateau-type models, while Higgs inflation may remain compatible once quantum-gravity
corrections are taken into account.

Therefore, in the remainder of this work, we quantify how LQC effective corrections modify the dynamics and the
predicted observables of non-minimally coupled Higgs inflation, and we assess whether the resulting $(n_s,r)$
remain compatible with current CMB constraints. These results will then be used to evaluate, within the
Liouville-measure framework introduced above, the probability of obtaining a sufficiently long inflationary phase.

During inflation one typically has $|\phi|\gg v_0$, so the symmetry-breaking scale is negligible. Accordingly, in the following, we set $v_0\simeq 0$.

\subsection{Cosmological perturbation}
\label{subsec:Cosmological perturbation}


For the non-minimal coupling \eqref{eq:non_minimal_coupling} and the Higgs potential \eqref{eq:higgs_potential},
the slow-roll quantities defined in \eqref{Slow_Para} take the form

\bea
&& \delta_{U}=-\frac{8 \xi  D_l^{-(n+1)}}{\kappa\left(1+ \xi  \phi^2\right)},\\
&& \delta_{X}=\frac{8 D_l^{m-2 (n+1)}}{\kappa  \phi ^2\left(1+  \xi  \phi ^2\right)}.
\eea
Substituting these expressions into \eqref{epsilon}, we obtain the first slow-roll parameter
\bea
&& \epsilon \simeq \frac{4 D_l^{-2 (n+1)} \left(2 D_l^m+\xi  \phi ^2 D_l^{n+1}\right)}{\kappa\phi ^2 \left(1+ \xi  \phi ^2\right)}.
\eea

The inflaton value at the end of inflation, $\phi_{f}$, is determined by the condition $\epsilon=1$, which yields
\be
\phi^2_{f}=\frac{\sqrt{\kappa ^2+24 \kappa  \xi +16 \xi ^2} - (\kappa -4 \xi) }{2 \kappa  \xi }.
\ee
At the level of the algebraic equation for $\phi_f^2$, there are two roots. We select the physically admissible
one with $\phi_f^2>0$.

We define another slow-roll parameter analogous to Eq. \eqref{eq:new_roll}
\be
\epsilon_s \equiv \frac{\kappa Q_{s}c_{s}^2}{U} \simeq \left(\delta_X\right)_{LQ},
\ee

Evaluated at horizon crossing
$(\phi,q)=(\phi_i,q_i)$, the scalar power spectrum is 
\be
\mathcal{P}_{s} = D_l^{m+2(n+1)}\frac{\kappa ^3 \lambda  \phi^6}{192 \pi ^2 \left(\xi  \phi^2+1\right)} \ \Bigg|_{\phi=\phi_i, \ q=q_i}.
\ee 
From this latter expression, we can solve the self-coupling parameter $\lambda(N_{*})$ as a function of the scalar field value at the horizon crossing $\phi_{*}=\phi(N_{*})$. Using the latest observational constraints, we consider the value $P_{s}=2.141\times 10^{-9}$ \cite{Planck:2018jri}.

So, the scalar spectral index and tensor-to-scalar ratio are

\be
\begin{split}
    n_s-1 &= \frac{D_l^{-\frac{m}{2}} \left[2 \phi ^2 \left(\kappa  (n+1) \delta _D \left(\xi  \phi ^2+1\right)-8 \xi \right)-8\right]-16 D_l^{-(n+1)}}{\kappa \phi ^2 \left(1+\xi  \phi ^2\right)} \ \Bigg|_{\phi=\phi_i, \ q=q_i}, \\
    r &= \frac{128 D_l^{m-2 (n+1)}}{\kappa  \phi ^2\left(1+  \xi  \phi ^2\right)}\Bigg|_{\phi=\phi_i, \ q=q_i}.
\end{split}
\ee

Finally, using eq. \eqref{def_running}, the running of the spectral index is computed as
\be
\begin{split}
    \alpha_s =\frac{d n_s}{dN}=\frac{64 \xi^2}{A_4^2 \left(A_4+\kappa \right){}^2} \left[D_l{}^{(n+1)-\frac{m}{2}}\left(2 \kappa A_4  +2 A_4^2+\kappa ^2\right) +2 \kappa  \left(2 A_4+\kappa \right)\right] \ \Bigg|_{q=q_i, \ N=N_i},
\end{split}
\ee
where $A_4$ is defined as

\be
A_4 \equiv D_l{}^{n+1} \xi  \left(\kappa  \phi _f^2+8 N\right).
\ee

For the quartic potential, the non-minimal coupling $\xi$ and the inverse-volume corrections act jointly to suppress $r$ and shift $n_s$ toward larger values, with the coupling $\lambda$ fixed at each $N$ by the power spectrum normalization $\mathcal{P}_s = 2.141\times 10^{-9}$ \cite{Planck:2018jri}.

\begin{figure}[H] 
    \centering
    \begin{minipage}{0.5\linewidth}
         \centering
    \includegraphics[width=0.8\textwidth]{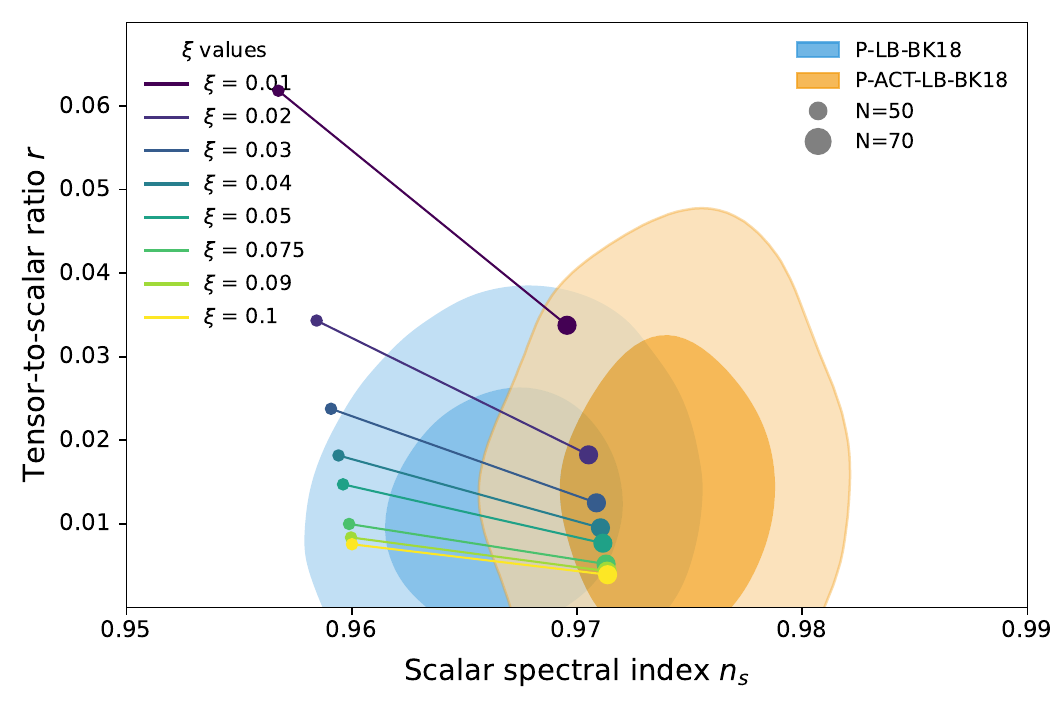}
    \end{minipage}\hfill
    \begin{minipage}{0.5\linewidth}
        \centering
        \includegraphics[width=0.8\textwidth]{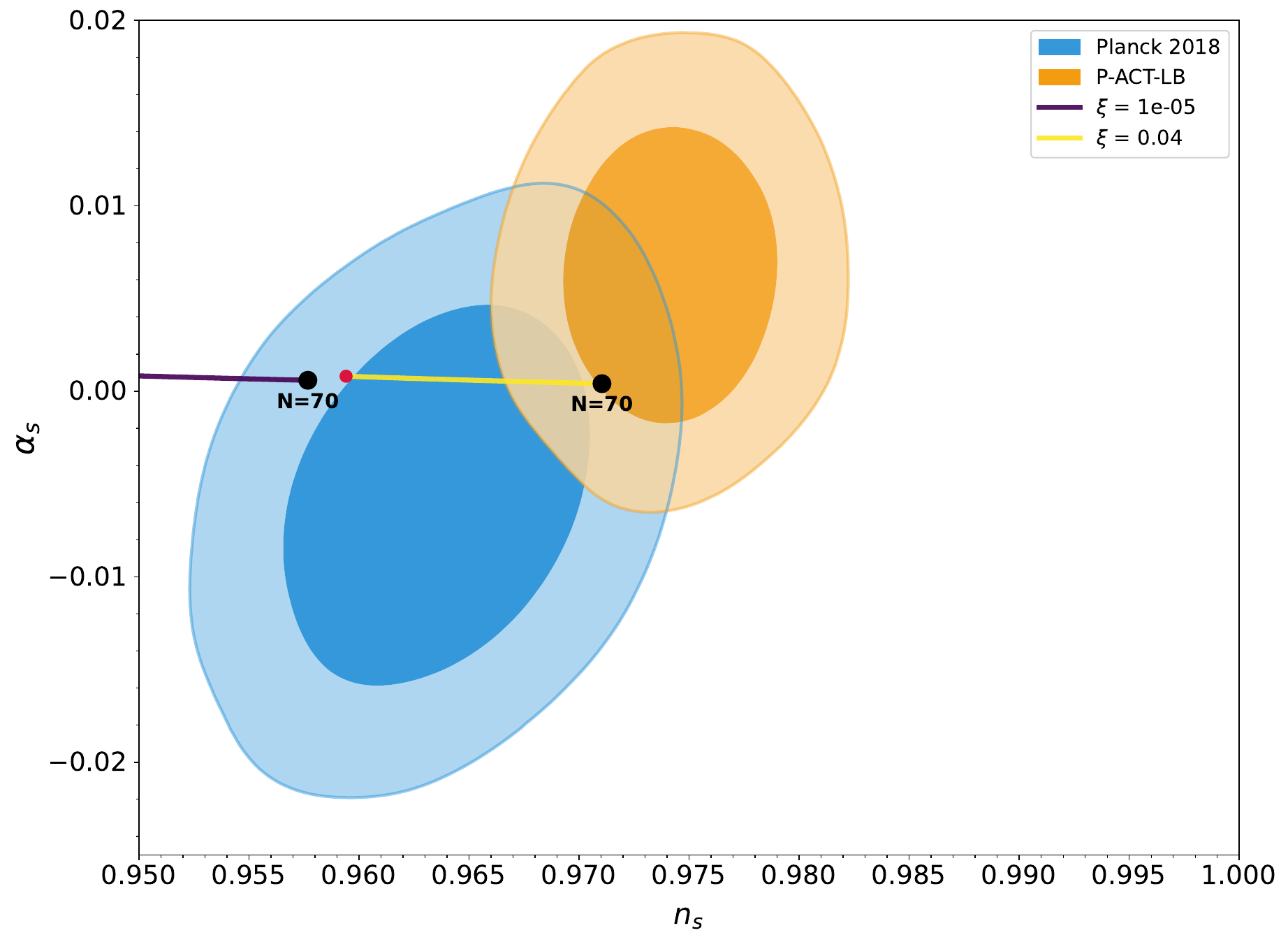}
    \end{minipage}
    \caption{Inflationary predictions for $V(\phi)\propto\phi^{4}$ in LQC with inverse-volume corrections. Left panel shows $r$ versus $n_s$ and right panel shows $\alpha_s$ versus $n_s$ for different non-minimal couplings $\xi$, with markers at $N=50$ and $N=70$. We fix the operator-ordering parameters  $m=0$, $n=0$, the inverse-volume eigenvalue parameter  $l=3/4$, and the initial value $q_i=100$. Here P-LB-BK18 denotes Planck 2018 + lensing + BICEP/Keck 2018 \cite{Planck:2018jri}, and P-ACT-LB-BK18 denotes Planck 2018 + ACT DR6 + lensing + BICEP/Keck 2018 \cite{Planck:2018jri, AtacamaCosmologyTelescope:2025blo, AtacamaCosmologyTelescope:2025nti}.}
    \label{fig:4_sp}
\end{figure}

\begin{figure}[H] 
    \centering
    \begin{minipage}{0.5\linewidth}
         \centering
    \includegraphics[width=0.8\textwidth]{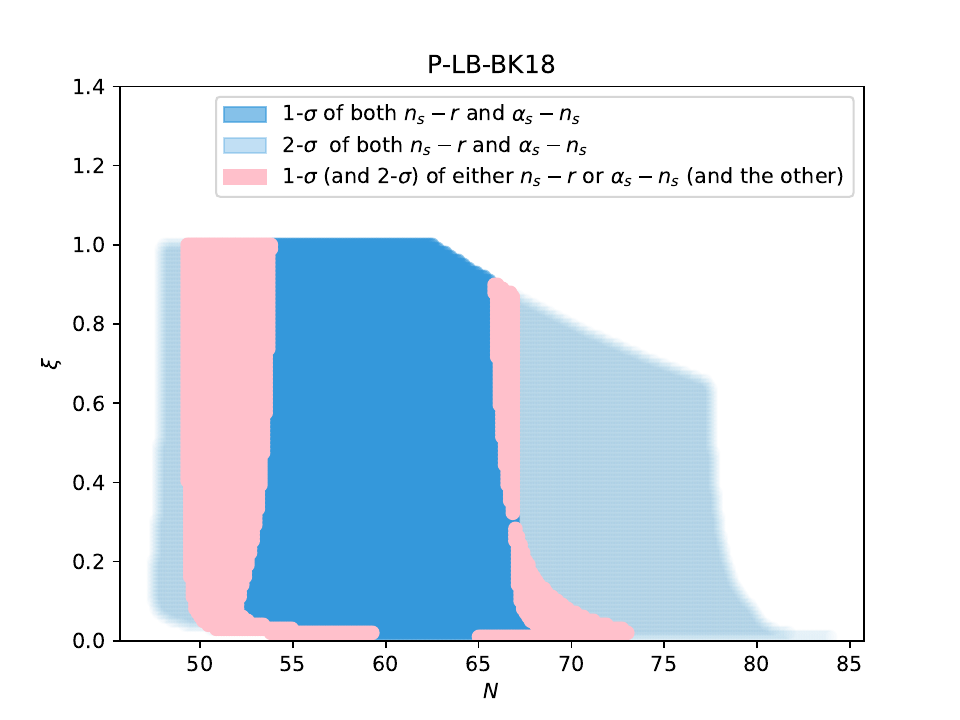}
    \end{minipage}\hfill
    \begin{minipage}{0.5\linewidth}
        \centering
        \includegraphics[width=0.8\textwidth]{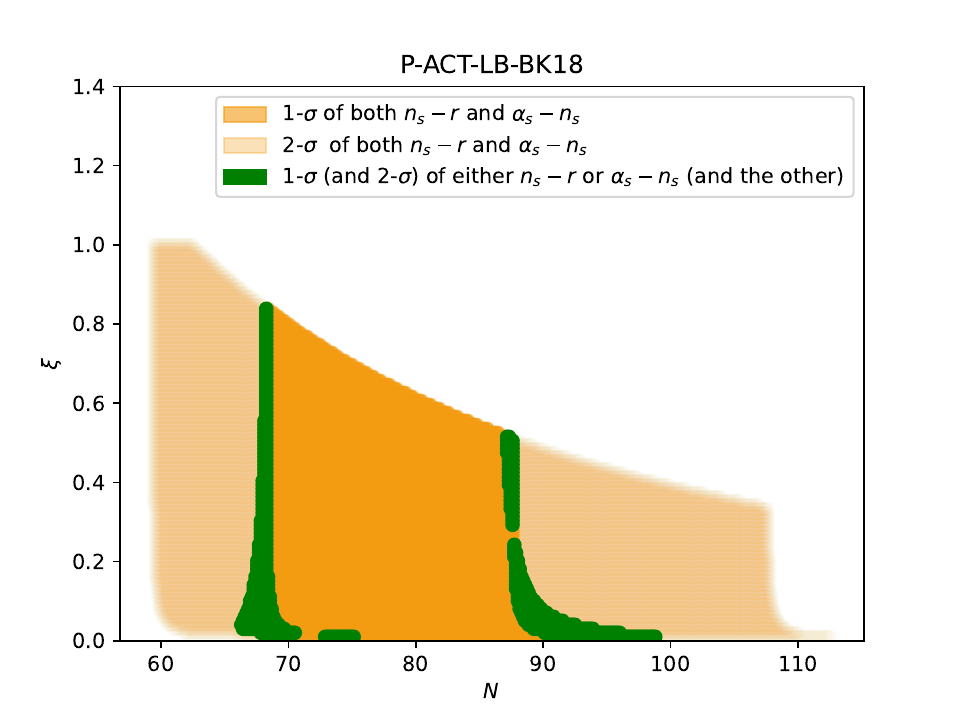}
    \end{minipage}
    \caption{Allowed region in the $(N,\xi)$ plane for the quartic potential $V(\phi)\propto\phi^{4}$ in LQC with inverse-volume corrections, imposing the theoretical bound $r\gtrsim 5 \times 10^{-4}$ \cite{Franciolini:2018ebs}. The left panel uses P-LB-BK18 (Planck 2018 + lensing + BICEP/Keck 2018 \cite{Planck:2018jri}) constraints and the right panel uses P-ACT-LB-BK18 (Planck 2018 + ACT DR6 + lensing + BICEP/Keck 2018 \cite{Planck:2018jri,AtacamaCosmologyTelescope:2025blo,AtacamaCosmologyTelescope:2025nti}) constraints. The three shaded regions correspond to different levels of joint observational compatibility: the innermost region satisfies $\leq 1\sigma$ in both the $(n_s,r)$ and $(n_s,\alpha_s)$ planes simultaneously; the outermost region satisfies $\leq 2\sigma$ in both planes simultaneously; and the intermediate region satisfies $\leq 1\sigma$ in one plane and $\leq 2\sigma$ in the other, representing a confidence level intermediate between the two. We fix the operator-ordering parameters $m=0$ and $n=0$, the inverse-volume eigenvalue parameter $l=3/4$, and the initial value $q_i=100$.}
    \label{fig:4_parameter}
\end{figure}

FIG.~\ref{fig:4_sp} summarizes our predictions for the quartic potential $V(\phi)\propto\phi^{4}$ in LQC, including inverse-volume corrections. The left panel shows the $(n_s,r)$ plane for several values of the non-minimal coupling $\xi$, while the right panel shows the corresponding trajectories in the $(n_s,\alpha_s)$ plane. The dots indicate the predictions evaluated at $N=50$ and $N=70$ e-folds. Throughout, we fix the operator-ordering parameters to $m=0$ and $n=0$, choose $l=3/4$, and set the initial value $q_i=100$. Our predictions are compared with the Planck 2018 and ACT+Planck+BICEP/Keck constraints, shown as shaded contours. FIG.~\ref{fig:4_parameter} complements this comparison by translating the observational bounds into constraints on the model parameters. Specifically, we project the viable solutions onto the $(N,\xi)$ plane and retain only those points that simultaneously satisfy the bounds in the $(n_s,r)$ and $(n_s,\alpha_s)$ planes. We also impose the lower limit $r\gtrsim 5\times10^{-4}$ \cite{Franciolini:2018ebs}. This representation makes explicit the allowed ranges of $\xi$ and the corresponding number of e-folds consistent with current data.

We now use these LQC-corrected observables to evaluate, within the Liouville-measure framework, the fraction of
initial conditions that yield sufficient inflation.
\subsection{Probability of inflation}
\label{subsec:Probability of inflation}

In the slow-roll regime, the loop-corrected Hubble parameter takes the form
\begin{equation}
    H^2=D_l^m\frac{\kappa\lambda\left(\phi^2-v_0^2\right)^2}{3\left(1+\xi\phi^2\right)}.
\end{equation}

The evolution of the inflaton in terms of the number of e-folds follows from Eq.~\eqref{eq:dPhi_dN}. For the present choice of potential and non-minimal coupling, we find
\begin{equation}\label{eq:dN_dPhi}
\frac{d\phi}{dN}=-4D_l^{(n+1)}\frac{\phi\left(1+\xi v_0^2\right)}{\kappa\left(\phi^2-v_0^2\right)},
\end{equation}
and thus\footnote{Here, we use integration by parts to pull $D_l$ out of the integral: $\int D_l f(\phi) d\phi = D_l \int f(\phi) d\phi - \int \delta_D D_l \frac{H}{\dot{\phi}} \left(\int f(\phi) d\phi \right) d\phi$. Since $\delta_D$ is a small first-order parameter in the slow-roll expansion, the second term is strictly higher-order. Therefore, to zeroth order, we can safely approximate $\int D_l f(\phi) d\phi \simeq D_l \int f(\phi) d\phi$.}
 \begin{equation}
 N = \frac{\kappa}{8\left(1+\xi v_0^2\right)}\left[2v_0^2\log{\left(\frac{\phi_{f}}{\left(\phi_{i}\right)^{D_l(q_{i})^{-(n+1)}}}\right)}+\left(D_l(q_{i})^{-(n+1)}\phi^2_{i}-\phi^2_{f}\right)\right].
\end{equation}

The momentum conjugate to $\phi$ is given by
\begin{equation}
    P_\phi = a_{\text{Pl}}^3 q^{3/2} \left(\dot{\phi} D_l^{-(n+1)}-\frac{6 H \xi  \phi }{\kappa }\right),
\end{equation}
and it is convenient to introduce the auxiliary quantity
    \begin{equation}
B_q=a_{\text{Pl}}^3 q^{3/2}\frac{ \left(4 D_l^m \kappa  \lambda  \phi  \left(\phi ^2-v_0^2\right) -6 H^2 \xi  \phi \right)}{3 H^2 \kappa }.
\end{equation}

The probability of realizing a sufficiently long inflationary phase is proportional to the corresponding Liouville measure $\mathcal{M}$ on the space of solutions. By using \eqref{eq:probability_final_exp}, we thus obtain 
\begin{equation}
\label{eq:probability}
    \begin{split}
        \mathcal{P} &\propto 2 a_{\text{Pl}}^3 q^{3/2}\sqrt{\frac{\lambda}{ 3\kappa}} \left[\sqrt{ \left(\xi  \phi _{i}^2+1\right)  \left(\phi ^2_{i}-v_0^2\right)^2 D_l(q_{i})^m}-\sqrt{\left(\xi  \phi_{f}^2+1\right)  \left(\phi ^2_{f}-v_0^2\right)^2}\right],
    \end{split}
\end{equation}
where $D_l(q_{f})\to 1$ is the classical limit.

\begin{figure}[H]
    \centering
    \begin{minipage}{0.5\linewidth}
        \centering
        \includegraphics[width=0.8\textwidth]{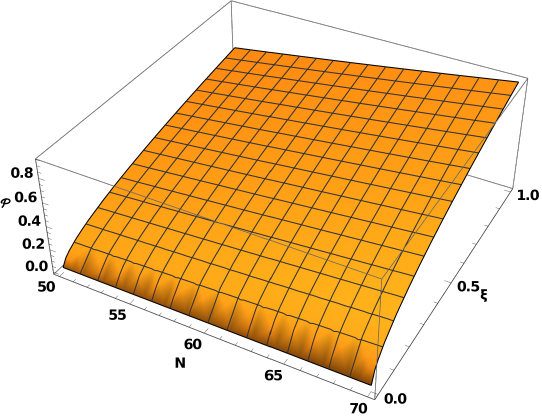}
    \end{minipage}\hfill
    \begin{minipage}{0.5\linewidth}
        \centering
        \includegraphics[width=0.8\textwidth]{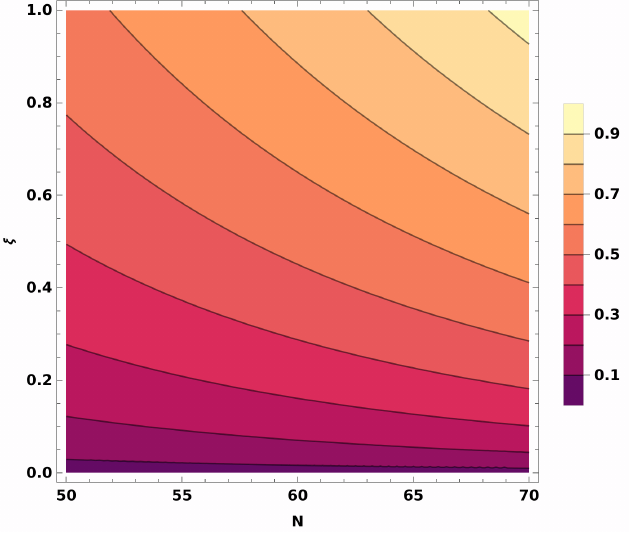}
    \end{minipage}
    \caption{Effect of the non-minimal coupling parameter $\xi$ on the probability of inflation $\mathcal{P}$ for the quartic potential $V(\phi)=\lambda \phi^{4}$, with $\lambda\simeq  6.14\times10^{-13}$ inferred from the scalar power-spectrum amplitude $\mathcal{P}_{s}=2.141\times 10^{-9}$ \cite{Planck:2018jri}. The left panel shows the surface $\mathcal{P}(N,\xi)$ and the right panel shows the corresponding contour map. We fix the inverse-volume eigenvalue parameter to $l=3/4$ and the initial value to $q_i=100$.}
    \label{fig:4_xi}
\end{figure}

To quantify the typicality of a successful slow-roll phase, we evaluate the Liouville measure of the effective phase space for the quartic potential. FIG.~\ref{fig:4_xi} shows how the probability of achieving a given number of e-folds, $N$, depends on the non-minimal coupling $\xi$, with the self-coupling $\lambda$ held fixed to correctly normalize the amplitude of the scalar power spectrum, $\mathcal{P}_{s}=2.141\times 10^{-9}$ \cite{Planck:2018jri}. Notably, for this Higgs-like potential, the probability of sufficient inflation exhibits a strong enhancement as the non-minimal coupling $\xi$ increases. The left panel provides a three-dimensional representation of the probability distribution, while the right panel shows the corresponding contour map. It is worth emphasizing that while these plots explore a broad theoretical parameter domain—specifically $\xi \in (0,1)$ and $N \in (0,70)$ in units where $M_{Pl}^2=1$—only a strictly bounded, highly localized subset of this space actually aligns with physical trajectories that satisfy current CMB bounds.

\section{String-inspired potentials with non-minimal coupling}
\label{sec:String-inspired potentials with non-minimal coupling}

In this section, we investigate string-motivated fractional monomial potentials in the presence of a non-minimal coupling to gravity. We adopt the coupling function given in Eq.~\eqref{eq:non_minimal_coupling} and focus on the two representative potentials
\begin{equation}
    V(\phi)=
    \begin{cases}
        \lambda \phi^{1/3}\\
        \lambda \phi^{2/3}
    \end{cases}.
\end{equation}
Monomial potentials of the form $V(\phi) \propto \phi^p$ with fractional powers $p < 1$ are strongly motivated by ultraviolet (UV) completions of inflation, particularly within the framework of string theory. For instance, axion monodromy inflation naturally generates such potentials, where the underlying shift symmetry of an axion field is broken by branes or fluxes, leading to a flattened potential structure. These ``flattened'' potentials are phenomenologically advantageous because they predict a significantly lower tensor-to-scalar ratio $r$ compared to standard chaotic inflation ($p=2, 4$), placing them comfortably within the precision bounds set by Planck and BICEP/Keck. Additionally, the recent trend in ACT DR6 observational data towards slightly larger $n_s$ values renews interest in these fractional power laws, as they occupy a favorable region of the $n_s-r$ plane that is distinct from the Starobinsky or Higgs attractors.

\subsection{Cosmological perturbation}
\label{subsec:Cosmological Perturbation}

\subsubsection{\texorpdfstring{$V(\phi)=\lambda \phi^{1/3}$}{V(phi) = lambda phi\textasciicircum(1/3)}}

For the fractional monomial potential $V(\phi)=\lambda\phi^{1/3}$, the loop-corrected slow-roll parameters take the form
\begin{equation}
    \begin{split}
         \delta_{X} &= \frac{\left(1-11 \xi  \phi ^2\right)^2 D_l^{m-2 (n+1)}}{18 \phi ^2 \left(\kappa  \xi  \phi ^2+\kappa \right)},\\
         \delta_{U} &= \frac{2 \xi  \left(11 \xi  \phi ^2-1\right) D_l^{-(n+1)}}{3 \left(\kappa  \xi  \phi ^2+\kappa \right)},\\
        \epsilon &= -\frac{\left(11 \xi  \phi ^2-1\right) D_l^{-2 (n+1)} \left(\left(1-11 \xi  \phi ^2\right) D_l^m+6 \xi  \phi ^2 D_l^{n+1}\right)}{18 \phi ^2 \left(\kappa  \xi  \phi ^2+\kappa \right)}.
    \end{split}
\end{equation}
Inflation ends when the slow-roll condition is violated, i.e. when $\epsilon\simeq 1$, which yields
\be
\phi^2_{f}=\frac{1}{3 \sqrt{9 \kappa ^2+18 \kappa  \xi +\xi ^2}+9 \kappa +8 \xi }.
\ee

The corresponding inflationary observables, evaluated at horizon exit ($\phi=\phi_i$, $q=q_i$), are
\be
\mathcal{P}_{s} = D_l^{m+2(n+1)}\frac{3 \kappa ^3 \lambda  \phi ^{7/3}}{4 \pi ^2 \left(1-11 \xi  \phi ^2\right)^2 \left(\xi  \phi ^2+1\right)} \ \Bigg|_{\phi=\phi_i, \ q=q_i},
\ee from which we obtain $\lambda(N_{*})$ at the horizon crossing, and 
\begin{equation}
    \begin{split}
        n_s-1 = & -\frac{2\left(1+ 13\xi  \phi ^2\right) D_l^{-\frac{m}{2}}}{3 \kappa \phi^2 \left(1+ \xi  \phi ^2\right)}+2 (n+1) \delta _D D_l^{-\frac{m}{2}}-\frac{\left(1-11 \xi  \phi ^2\right)^2 D_l^{-(n+1)}}{9 \kappa \phi ^2 \left( 1+ \xi  \phi ^2 \right)}\Bigg|_{\phi=\phi_i, \ q=q_i},\\
        r = & \frac{8 \left(1-11 \xi  \phi ^2\right)^2 D_l^{m-2 (n+1)}}{9 \kappa \phi ^2\left(1+ \xi  \phi ^2\right)}\Bigg|_{\phi=\phi_i, \ q=q_i}.
    \end{split}
\end{equation}

Also, by using Eq. \eqref{def_running}, the running of the spectral index is expressed as
\be
\begin{split}
    \alpha_s = & \frac{242 B_{\frac{1}{3}} \xi ^2 A_{\frac{1}{3}}^{D_l^{-(n+1)}} D_l^{-\frac{m}{2}-2 (n+1)}}{27 \kappa ^2 \left(-13 B_{\frac{1}{3}} A_{\frac{1}{3}}^{D_l^{-(n+1)}}+A_{\frac{1}{3}}^{2 D_l^{-(n+1)}}+12 B_{\frac{1}{3}}^2\right)^2} \\
    &\left[24 B_{\frac{1}{3}} A_{\frac{1}{3}}^{D_l^{-(n+1)}} \left(11 D_l^{m/2}-12 D_l^{n+1}\right)+A_{\frac{1}{3}}^{2 D_l^{-(n+1)}} \left(78 D_l^{n+1}-143 D_l^{m/2}\right)+936 B_{\frac{1}{3}}^2 D_l^{n+1}\right] \ \Bigg|_{q=q_i, \ N=N_i},
\end{split}
\ee

with

\be
A_{\frac{1}{3}} \equiv 1-11 \xi  \phi _f^2, \qquad B_{\frac{1}{3}} \equiv e^{\frac{22 \xi  N D_l^{-(n+1)}}{3 \kappa }}.
\ee

Thus, for $V(\phi)\propto\phi^{1/3}$, the fractional power of the potential already predicts a lower tensor-to-scalar ratio compared to the quartic case, and the non-minimal coupling $\xi$ provides an additional tuning of $n_s$ and $r$ toward the observationally favored region.

\begin{figure}[H] 
    \centering
    \begin{minipage}{0.5\linewidth}
         \centering
    \includegraphics[width=0.8\textwidth]{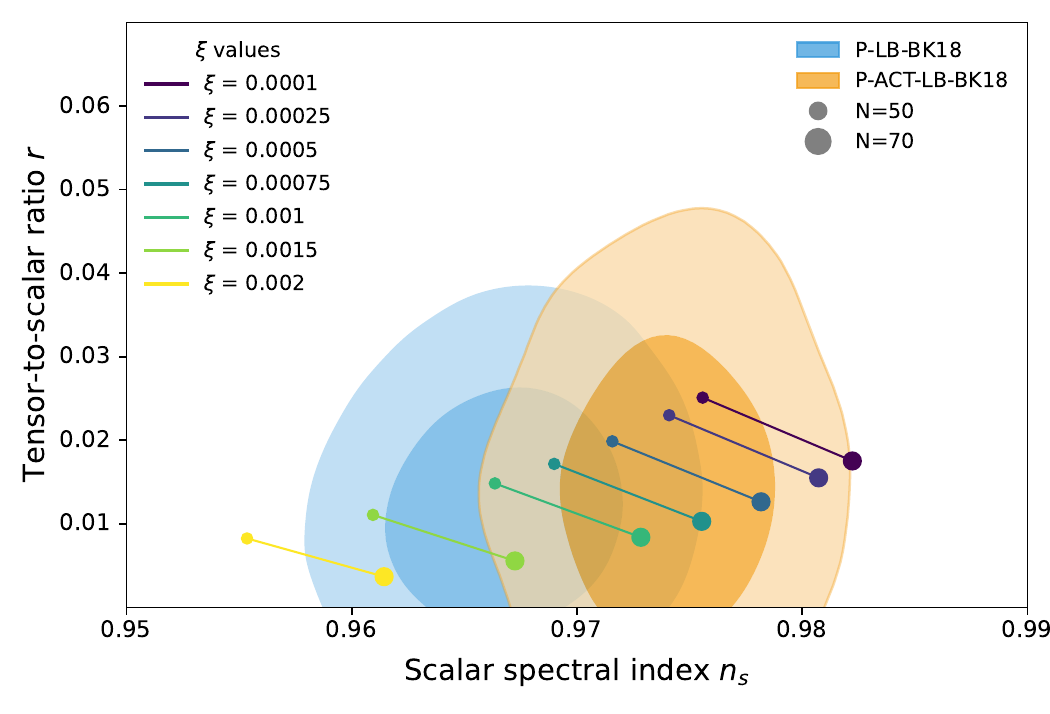}
    \end{minipage}\hfill
    \begin{minipage}{0.5\linewidth}
        \centering
        \includegraphics[width=0.8\textwidth]{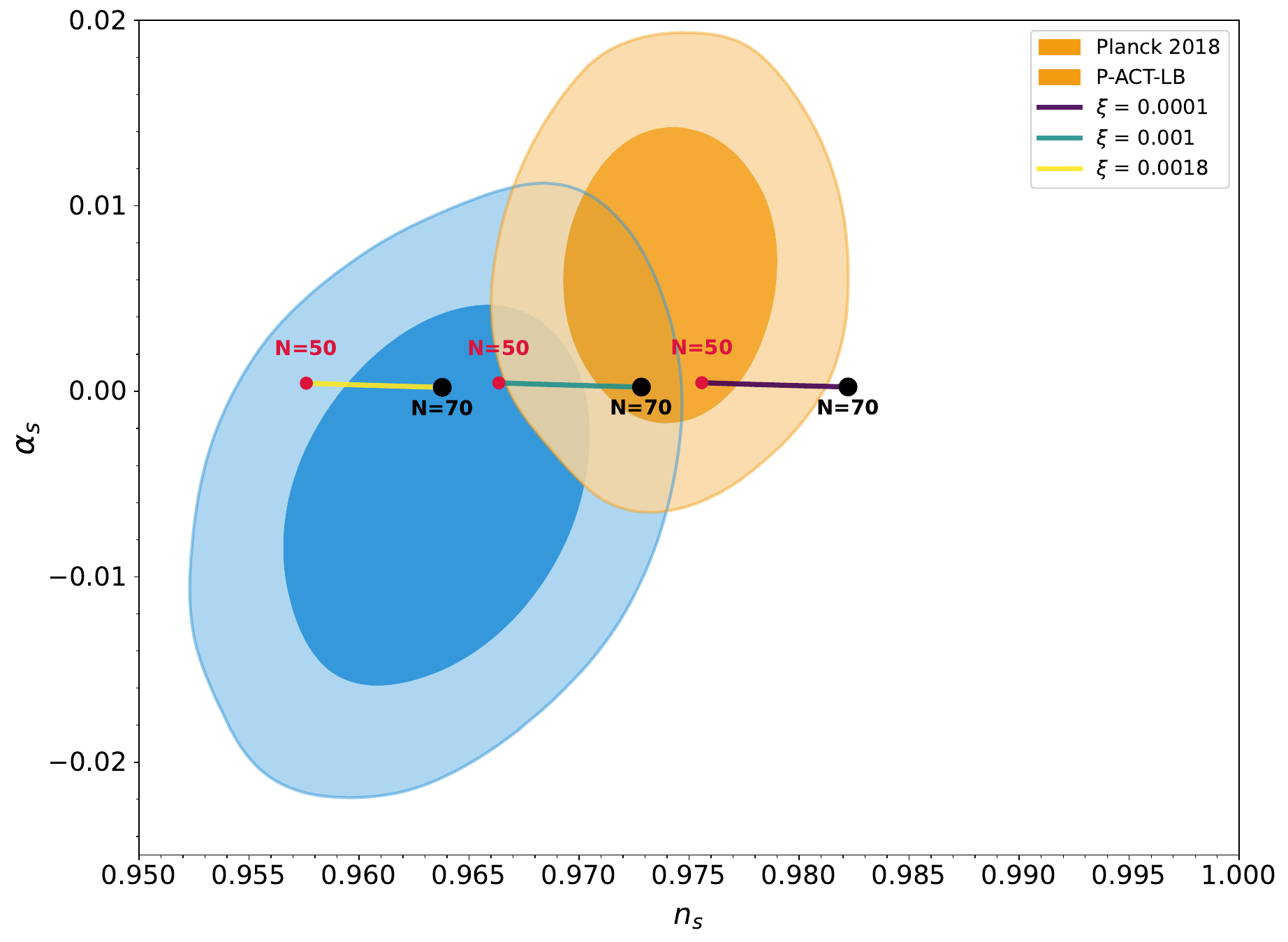}
    \end{minipage}
    \caption{Inflationary predictions for $V(\phi)\propto\phi^{1/3}$ in LQC with inverse-volume corrections. Left panel shows $r$ versus $n_s$ and right panel shows $\alpha_s$ versus $n_s$ for different non-minimal couplings $\xi$, with markers at $N=50$ and $N=70$. We fix the operator-ordering parameters  $m=0$, $n=0$, the inverse-volume eigenvalue parameter  $l=3/4$, and the initial value $q_i=100$. Here P-LB-BK18 denotes Planck 2018 + lensing + BICEP/Keck 2018 \cite{Planck:2018jri}, and P-ACT-LB-BK18 denotes Planck 2018 + ACT DR6 + lensing + BICEP/Keck 2018 \cite{Planck:2018jri, AtacamaCosmologyTelescope:2025blo, AtacamaCosmologyTelescope:2025nti}.}
    \label{fig:1_3_sp}
\end{figure}

\begin{figure}[H] 
    \centering
    \begin{minipage}{0.5\linewidth}
         \centering
    \includegraphics[width=0.8\textwidth]{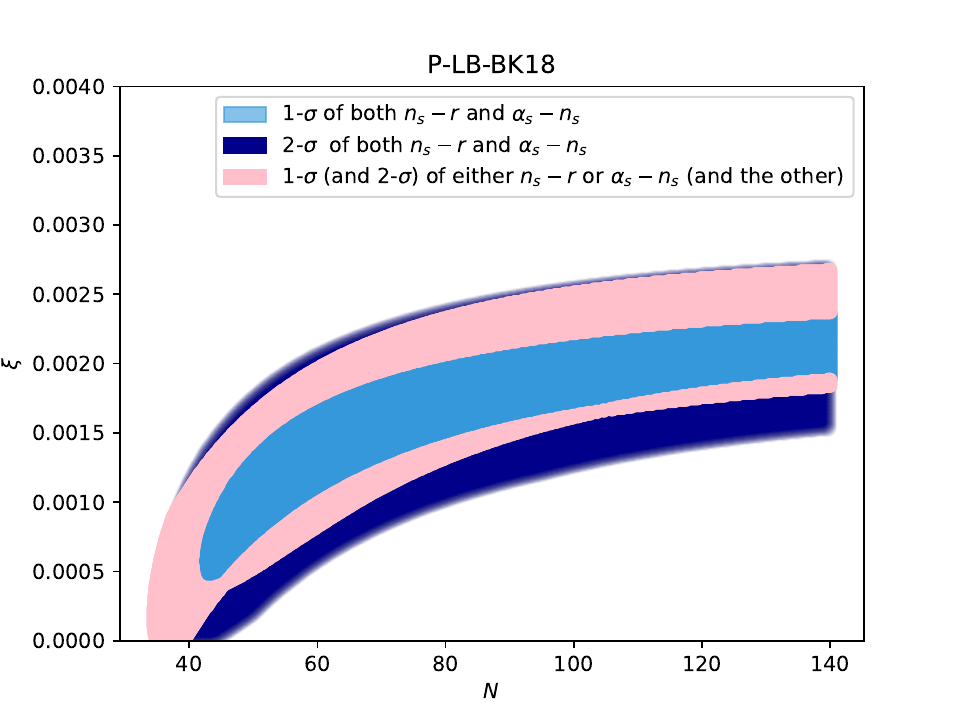}
    \end{minipage}\hfill
    \begin{minipage}{0.5\linewidth}
        \centering
        \includegraphics[width=0.8\textwidth]{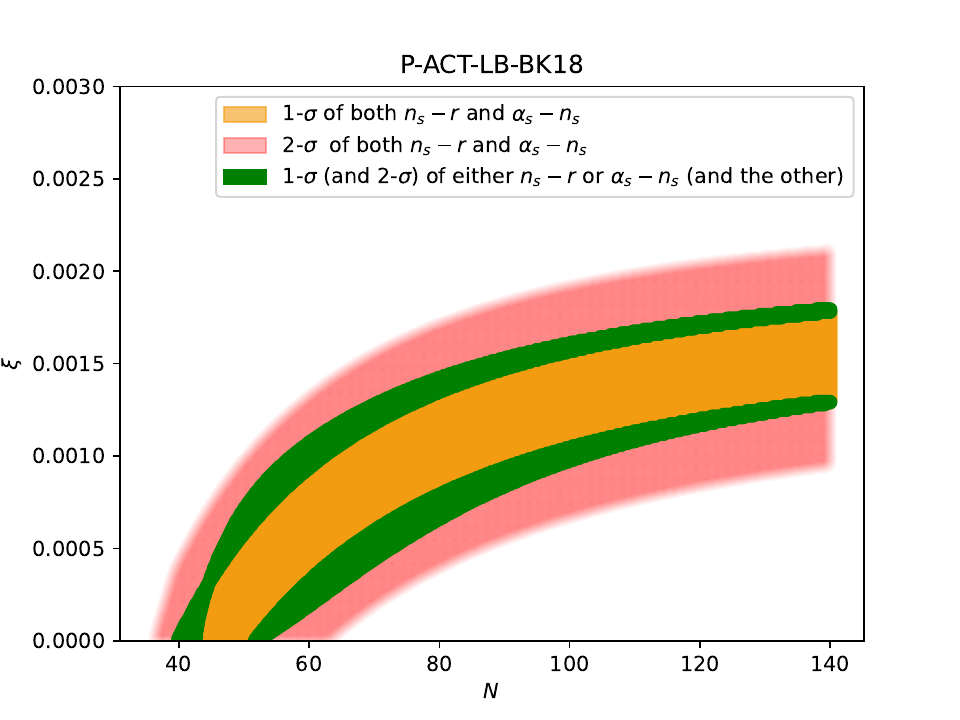}
    \end{minipage}
    \caption{Allowed region in the $(N,\xi)$ plane for the quartic potential $V(\phi)\propto\phi^{1/3}$ in LQC with inverse-volume corrections. The left panel uses P-LB-BK18 (Planck 2018 + lensing + BICEP/Keck 2018 \cite{Planck:2018jri}) constraints and the right panel uses P-ACT-LB-BK18 (Planck 2018 + ACT DR6 + lensing + BICEP/Keck 2018 \cite{Planck:2018jri,AtacamaCosmologyTelescope:2025blo,AtacamaCosmologyTelescope:2025nti}) constraints. The three shaded regions correspond to different levels of joint observational compatibility: the innermost region satisfies $\leq 1\sigma$ in both the $(n_s,r)$ and $(n_s,\alpha_s)$ planes simultaneously; the outermost region satisfies $\leq 2\sigma$ in both planes simultaneously; and the intermediate region satisfies $\leq 1\sigma$ in one plane and $\leq 2\sigma$ in the other, representing a confidence level intermediate between the two. We fix the operator-ordering parameters $m=0$ and $n=0$, the inverse-volume eigenvalue parameter $l=3/4$, and the initial value $q_i=100$.}
    \label{fig:1_3_parameter}
\end{figure}

In FIG.~\ref{fig:1_3_sp}, we present the theoretical predictions for the fractional monomial potential $V(\phi) \propto \phi^{1/3}$ under the influence of inverse-volume corrections. The left and right panels display the $(n_s, r)$ and $(n_s, \alpha_s)$ planes, respectively, for various choices of the non-minimal coupling $\xi$. As before, we set the ambiguity parameters to $m=0$ and $n=0$, with $l=3/4$ and an initial value $q_i=100$. The theoretical curves, evaluated for an e-folding range between $N=50$ and $N=70$, are superimposed on the $1\sigma$ and $2\sigma$ confidence contours from Planck 2018 and the combined ACT data. To better understand the viable parameter space, FIG.~\ref{fig:1_3_parameter} maps these observational bounds directly onto the $(N, \xi)$ plane. This projection isolates the specific combinations of coupling strength and inflationary duration that successfully reproduce the observed CMB power spectra.

\subsubsection{\texorpdfstring{$V(\phi)=\lambda \phi^{2/3}$}{V(phi) = lambda phi\textasciicircum(2/3)}}

For the fractional monomial potential $V(\phi)=\lambda \phi^{2/3}$, we get the loop-corrected slow-roll parameters
\begin{equation}
    \begin{split}
         \delta_{X} &=\frac{2 \left(1-5 \xi  \phi ^2\right)^2 D_l^{m-2 (n+1)}}{9 \kappa \phi ^2 \left(1+ \xi  \phi ^2 \right)},\\
         \delta_{U} &= \frac{4 \xi  \left(5 \xi  \phi ^2-1\right) D_l^{-(n+1)}}{3\kappa \left(1+ \xi  \phi ^2\right)},\\
        \epsilon &= -\frac{2 \left(5 \xi  \phi ^2-1\right) D_l^{-2 (n+1)} \left[\left(1-5 \xi  \phi ^2\right) D_l^m+3 \xi  \phi ^2 D_l^{n+1}\right]}{9 \kappa\phi ^2 \left(1+\xi  \phi ^2\right)}.
    \end{split}
\end{equation}

From the condition $\epsilon=1$ at the end of inflation, we have the final value of the inflaton
\be
\phi^2_{f}=\frac{4}{3 \sqrt{9 \kappa ^2+36 \kappa  \xi +4 \xi ^2}+9 \kappa +14 \xi }.
\ee

The scalar power spectrum at the horizon crossing becomes
\be
  \mathcal{P}_{s} = D_l^{m+2(n+1)}\frac{3 \kappa ^3 \lambda  \phi ^{8/3}}{16 \pi ^2 \left(1-5 \xi  \phi ^2\right)^2 \left(\xi  \phi ^2+1\right)} \ \Bigg|_{\phi=\phi_i, \ q=q_i},
  \ee which allows us to calculate $\lambda(N_{*})$.
Also, the corresponding spectral index and tensor-to-scalar ratio, evaluated at horizon exit ($\phi=\phi_i$, $q=q_i$), are given by
\begin{equation}
    \begin{split}
        n_s-1 = & \frac{2 \left[D_l^{-\frac{m}{2}} \left(3 \phi ^2 \left(3 \kappa  (n+1) \delta _D \left(\xi  \phi ^2+1\right)-14 \xi \right)-6\right)-2 \left(1-5 \xi  \phi ^2\right)^2 D_l^{-(n+1)}\right]}{9 \kappa \phi ^2 \left( 1+ \xi  \phi ^2 \right)}\Bigg|_{\phi=\phi_i, \ q=q_i},\\
        r = & \frac{32 \left(1-5 \xi  \phi ^2\right)^2 D_l^{m-2 (n+1)}}{9 \kappa\phi ^2 \left( 1+ \xi  \phi ^2 \right)}\Bigg|_{\phi=\phi_i, \ q=q_i},
        \end{split}
        \end{equation}
and the running of the scalar spectral index yields
{\small
\bea
\alpha_s=  \frac{400 B_{\frac{2}{3}} \xi ^2 A_{\frac{2}{3}}^{D_l^{-(n+1)}} D_l^{-\frac{m}{2}-(n+1)}}{27 \kappa ^2 \left[B_{\frac{2}{3}} A_{\frac{2}{3}}^{D_l^{-(n+1)}} \left(B_{\frac{2}{3}} A_{\frac{2}{3}}^{D_l^{-(n+1)}}-7\right)+6\right]^2} \Bigg[ -12 B_{\frac{2}{3}} A_{\frac{2}{3}}^{D_l^{-n-1}} \left(5 D_l^{m/2}-6 D_l^{n+1}\right)\nonumber\\
      +7 B_{\frac{2}{3}}^2 A_{\frac{2}{3}}^{2 D_l^{-(n+1)}} \left(5 D_l^{m/2}-3 D_l^{n+1}\right)-126 D_l^{n+1} \Bigg]\Bigg|_{q=q_i,N=N_i},
        \eea}
where we defined
\be
A_{\frac{2}{3}} \equiv 1-5 \xi  \phi _f^2, \qquad B_{\frac{2}{3}} \equiv e^{\frac{20 \xi  N D_l^{-(n+1)}}{3 \kappa }}.
\ee

In summary, the $V(\phi)\propto\phi^{2/3}$ potential yields predictions qualitatively similar to the $p=1/3$ case, with the non-minimal coupling $\xi$ again providing the primary handle for tuning the observables into agreement with current CMB constraints.

\begin{figure}[H] 
    \centering
    \begin{minipage}{0.5\linewidth}
         \centering
    \includegraphics[width=0.8\textwidth]{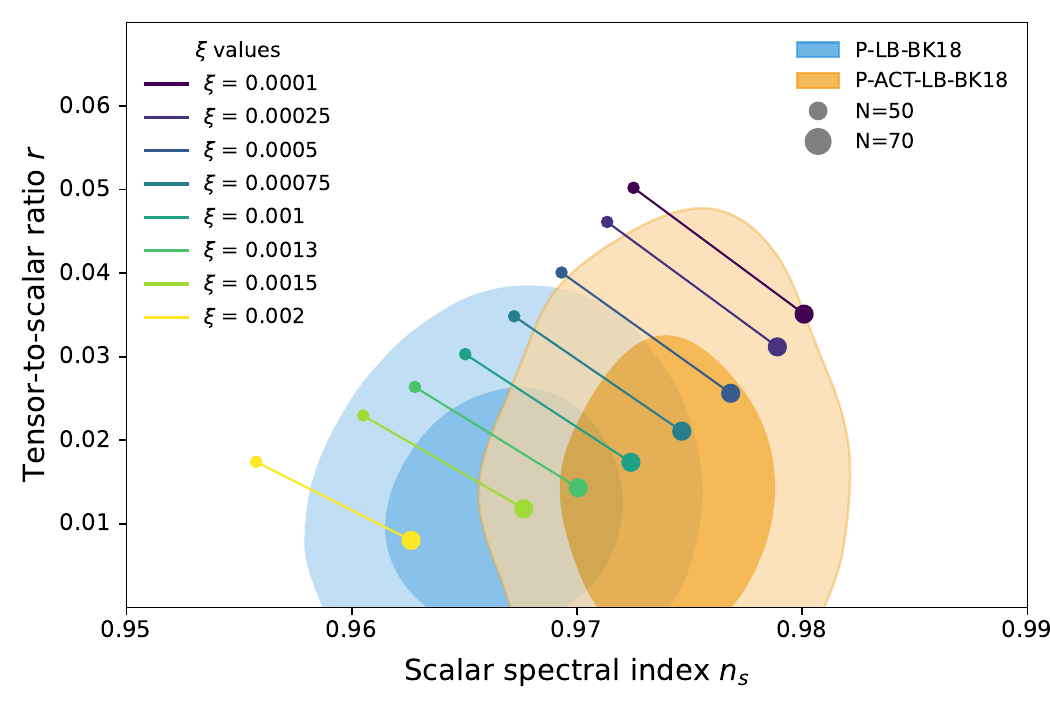}
    \end{minipage}\hfill
    \begin{minipage}{0.5\linewidth}
        \centering
        \includegraphics[width=0.8\textwidth]{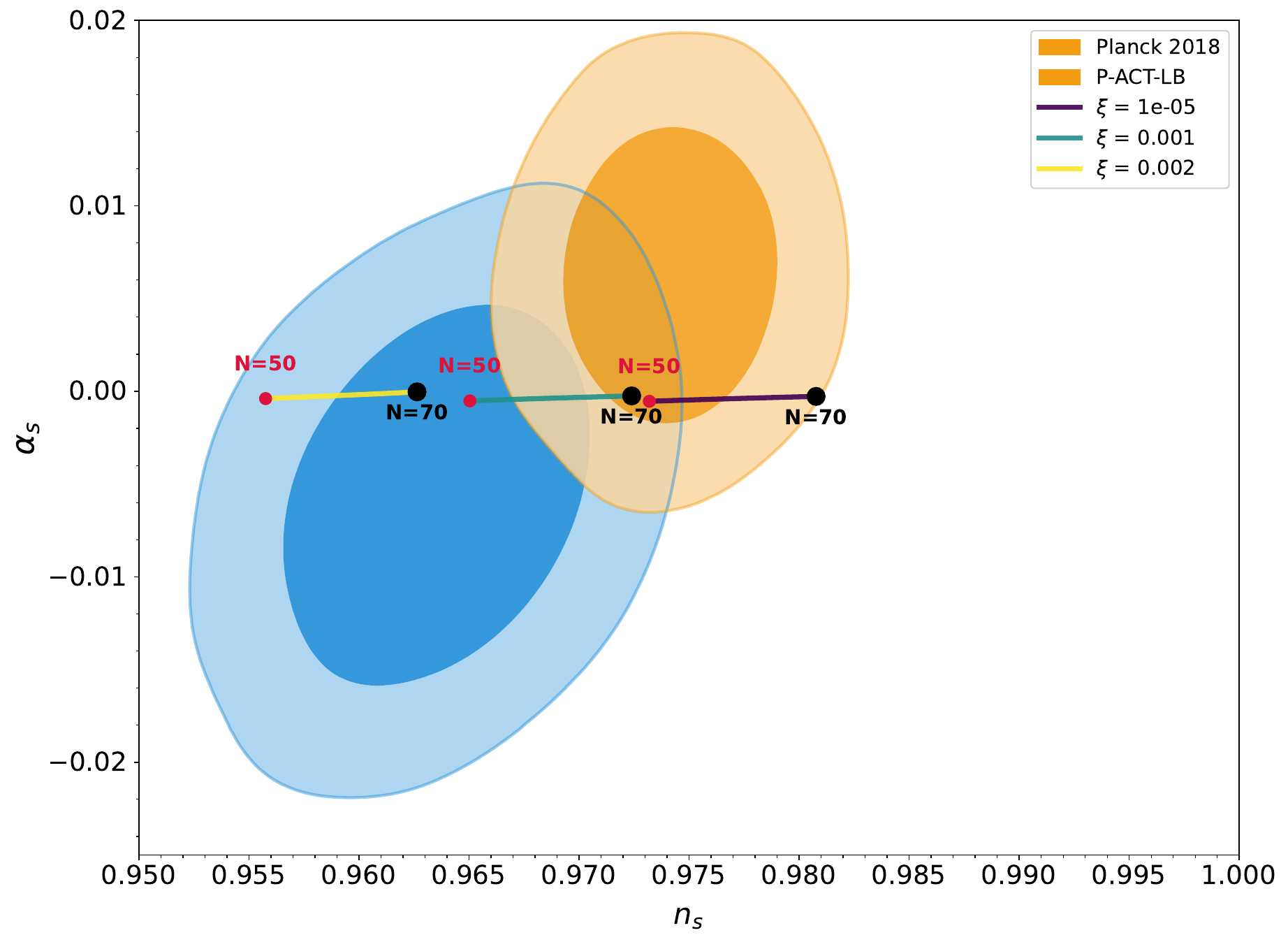}
    \end{minipage}
    \caption{Inflationary predictions for $V(\phi)\propto\phi^{2/3}$ in LQC with inverse-volume corrections. Left panel shows $r$ versus $n_s$ and right panel shows $\alpha_s$ versus $n_s$ for different non-minimal couplings $\xi$, with markers at $N=50$ and $N=70$. We fix the operator-ordering parameters  $m=0$, $n=0$, the inverse-volume eigenvalue parameter  $l=3/4$, and the initial value $q_i=100$. Here P-LB-BK18 denotes Planck 2018 + lensing + BICEP/Keck 2018 \cite{Planck:2018jri}, and P-ACT-LB-BK18 denotes Planck 2018 + ACT DR6 + lensing + BICEP/Keck 2018 \cite{Planck:2018jri, AtacamaCosmologyTelescope:2025blo, AtacamaCosmologyTelescope:2025nti}.}
    \label{fig:2_3_sp}
\end{figure}

\begin{figure}[H] 
    \centering
    \begin{minipage}{0.5\linewidth}
         \centering
    \includegraphics[width=0.8\textwidth]{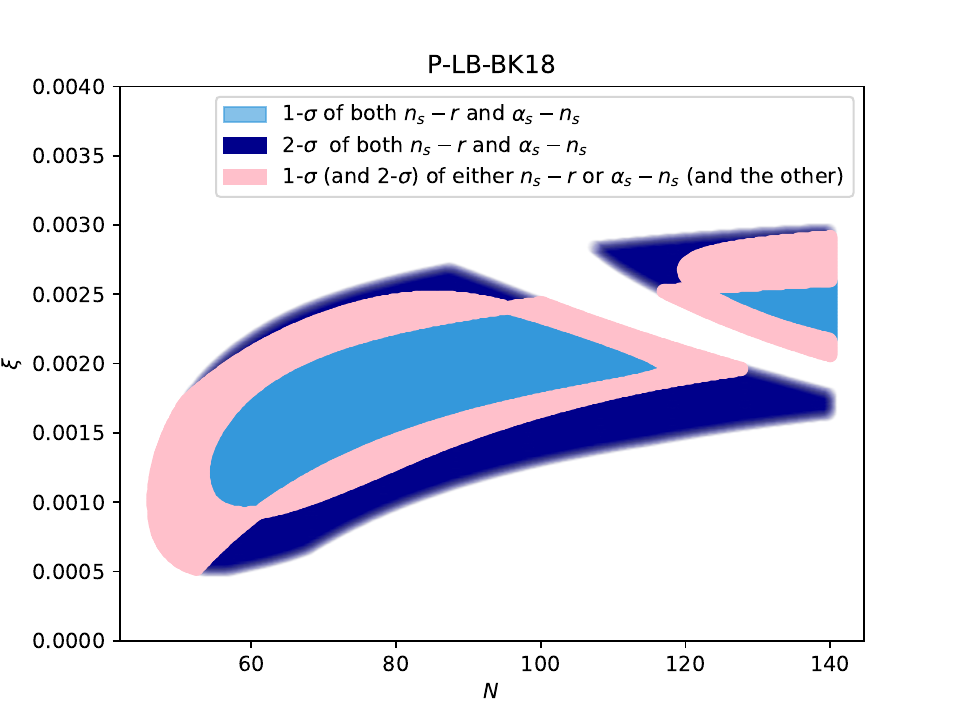}
    \end{minipage}\hfill
    \begin{minipage}{0.5\linewidth}
        \centering
        \includegraphics[width=0.8\textwidth]{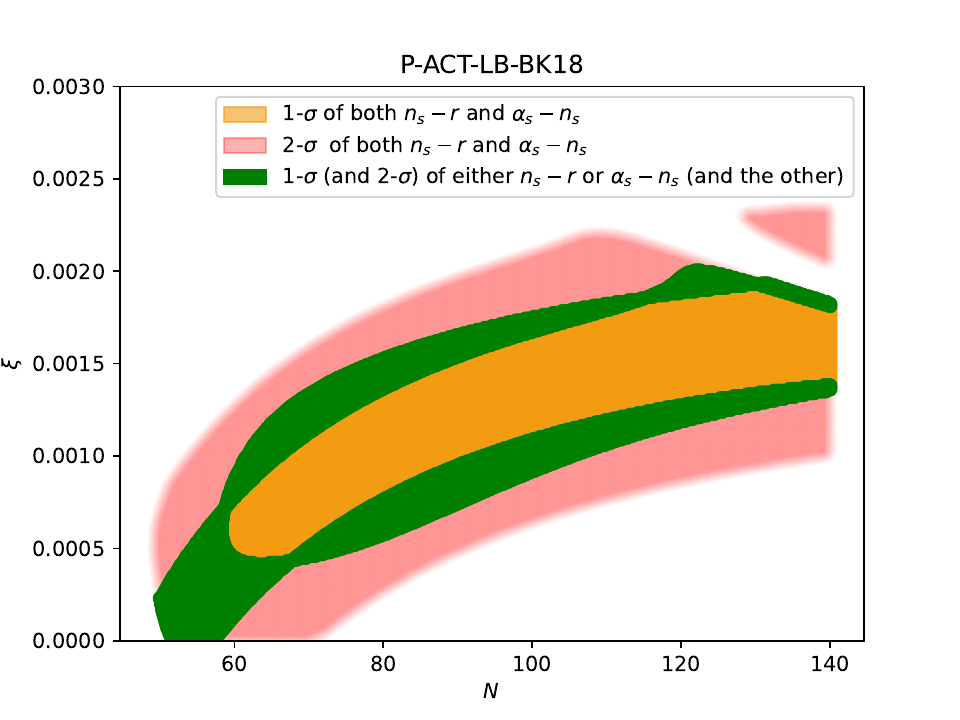}
    \end{minipage}
    \caption{Allowed region in the $(N,\xi)$ plane for the quartic potential $V(\phi)\propto\phi^{2/3}$ in LQC with inverse-volume corrections. The left panel uses P-LB-BK18 (Planck 2018 + lensing + BICEP/Keck 2018 \cite{Planck:2018jri}) constraints and the right panel uses P-ACT-LB-BK18 (Planck 2018 + ACT DR6 + lensing + BICEP/Keck 2018 \cite{Planck:2018jri,AtacamaCosmologyTelescope:2025blo,AtacamaCosmologyTelescope:2025nti}) constraints. The three shaded regions correspond to different levels of joint observational compatibility: the innermost region satisfies $\leq 1\sigma$ in both the $(n_s,r)$ and $(n_s,\alpha_s)$ planes simultaneously; the outermost region satisfies $\leq 2\sigma$ in both planes simultaneously; and the intermediate region satisfies $\leq 1\sigma$ in one plane and $\leq 2\sigma$ in the other, representing a confidence level intermediate between the two. We fix the operator-ordering parameters $m=0$ and $n=0$, the inverse-volume eigenvalue parameter $l=3/4$, and the initial value $q_i=100$.}
    \label{fig:2_3_parameter}
\end{figure}

The corresponding results for the $V(\phi) \propto \phi^{2/3}$ potential are illustrated in FIG.~\ref{fig:2_3_sp}. Here again, we plot the predicted pairs of $(n_s, r)$ and $(n_s, \alpha_s)$ against the background constraints, maintaining the same LQC parameter choices ($m=n=0$, $l=3/4$, and $q_i=100$). The markers on the curves denote the standard window of 50 to 70 e-folds. Following the same procedure as in the previous cases, we then constrain the underlying model parameters by filtering the allowed points. The outcome is displayed in FIG.~\ref{fig:2_3_parameter}, which highlights the region in the $(N, \xi)$ parameter space where the model remains fully consistent with both the scalar tilt and the bounds on the running and tensor modes.

\begin{figure}[H] 
    \centering
    \begin{minipage}{0.5\linewidth}
        \centering
        \includegraphics[width=0.8\textwidth]{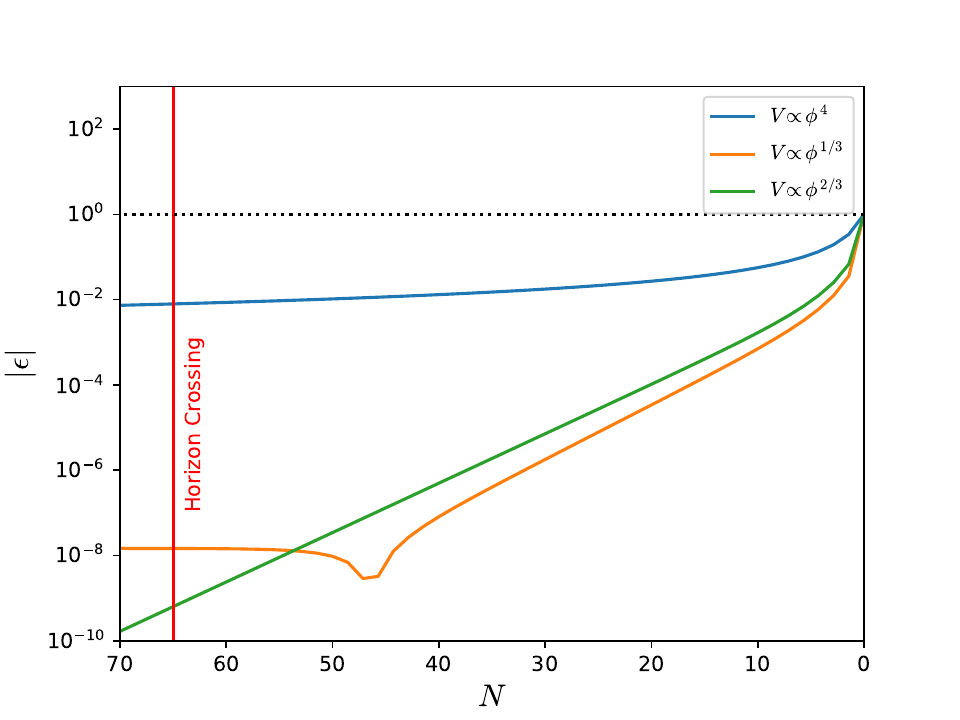}
    \end{minipage}\hfill
    \begin{minipage}{0.5\linewidth}
        \centering
        \includegraphics[width=0.8\textwidth]{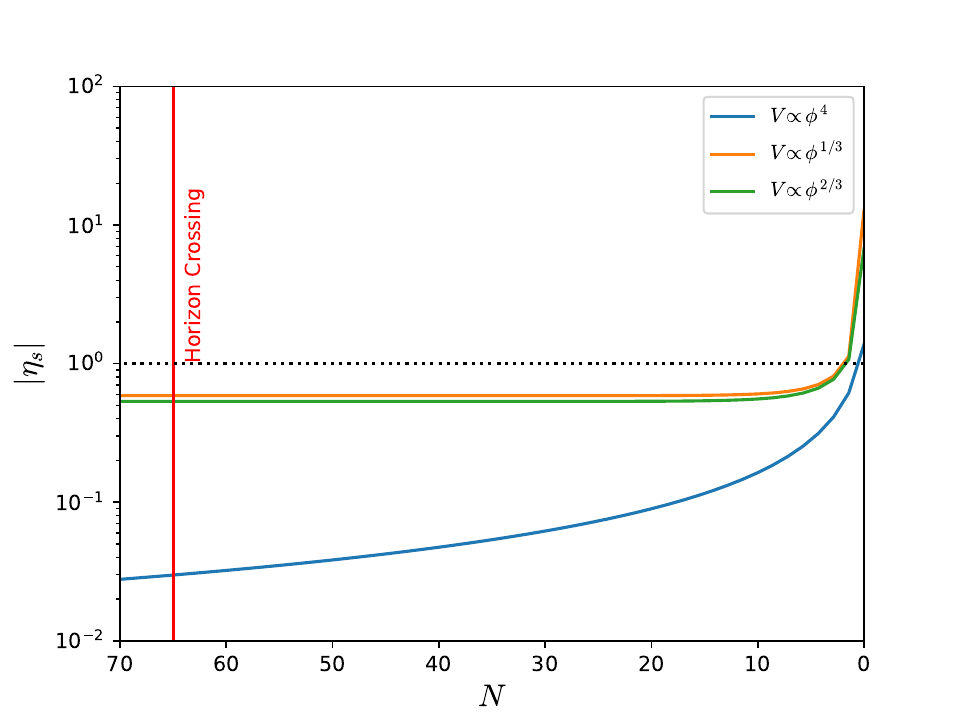}
    \end{minipage}
    \caption{Evolution of the LQC-modified slow-roll parameters $|\epsilon_{LQ}|$ (left panel) and $|(\eta_s)_{LQ}|$ (right panel) as functions of the number of e-folds $N$, shown for all three potentials considered in this work: $V \propto \phi^4$ (blue), $V \propto \phi^{1/3}$ (orange), and $V \propto \phi^{2/3}$ (green). Results are obtained with $l=3/4$, operator-ordering parameters $m=0$, $n=0$, and initial value $q_i=100$. The horizontal dotted lines mark the threshold values $\epsilon_{LQ}=1$ and $(\eta_s)_{LQ}=1$, signaling the end of slow-roll inflation. The red vertical line indicates horizon crossing. Deep in the inflationary phase, both parameters remain well below unity, confirming the self-consistency of the slow-roll approximation throughout the observable e-folding window.}
    \label{fig:slow_roll_evolution}
\end{figure}

To explicitly verify the internal consistency of our approximation, we must track the dynamical evolution of the LQC-modified slow-roll parameters throughout the expansion. The left and right panels of FIG.~\ref{fig:slow_roll_evolution} illustrate the behavior of $|\epsilon_{LQ}|$ and $|(\eta_s)_{LQ}|$, respectively, as functions of the e-folding number $N$. Deep in the inflationary regime, both parameters remain safely below unity, which guarantees a sustained quasi-de Sitter expansion. As the scalar field rolls toward the minimum of its effective potential, its kinetic energy gradually increases until the slow-roll conditions are no longer satisfied. This natural termination of slow-roll inflation is marked in the plots by the horizontal dotted lines at $\epsilon_{LQ}=1$ and $(\eta_s)_{LQ}=1$.

\subsection{Probability of inflation}
\label{subsec:Probability of inflation2}

\subsubsection{\texorpdfstring{$V(\phi)=\lambda \phi^{1/3}$}{V(phi) = lambda phi\textasciicircum(1/3)}}

For this potential, under the slow-roll approximation, one finds
\begin{equation}
\frac{d\phi}{dN} = \frac{\left(11 \xi  \phi^2-1\right) D_l^{n+1}}{3 \kappa  \phi}.
\end{equation}
Thus, the number of $e$-folds is computed as 
\begin{equation}
N= \frac{3 \kappa  \log \left(11 \xi  \phi ^2-1\right) D_l^{-(n+1)}}{22 \xi }\Bigg|_{\phi=\phi_{i}}^{\phi=\phi_{f}}.
\end{equation}

We obtain the probability of inflation to be
\begin{equation}
    \begin{split}
        \mathcal{P} &\propto 2 a_{\text{Pl}}^3 q^{3/2}\sqrt{\frac{\lambda}{ 3\kappa}} \left[\sqrt{ \left(\xi  \phi _{i}^2+1\right)  \phi^{1/3}_{i} D_l(q_{i})^m}-\sqrt{\left(\xi  \phi_{f}^2+1\right)  \phi^{1/3}_{f}}\right],
    \end{split}
\end{equation}
where we have considered $D_l(q_{f})=1$ .

\begin{figure}[H]
    \centering
    \begin{minipage}{0.5\linewidth}
        \centering
        \includegraphics[width=0.8\textwidth]{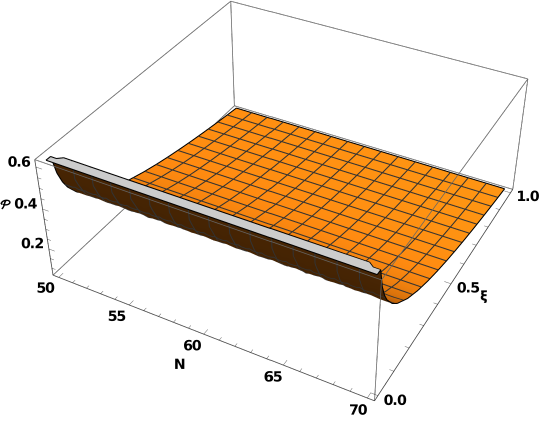}
    \end{minipage}\hfill
    \begin{minipage}{0.5\linewidth}
        \centering
        \includegraphics[width=0.8\textwidth]{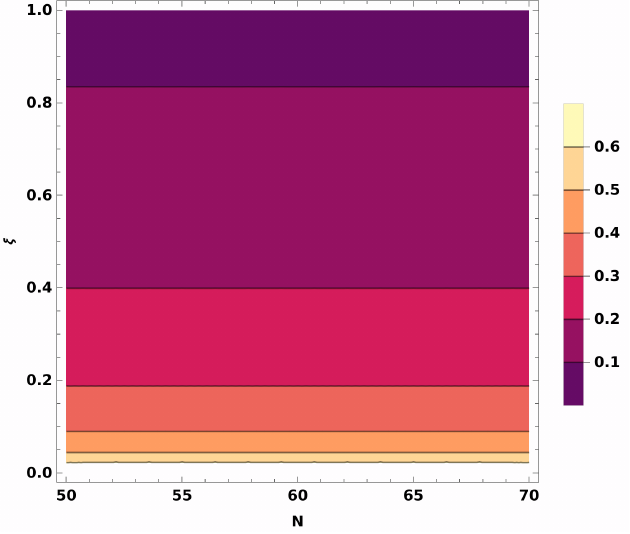}
    \end{minipage}
    \caption{Effect of the non-minimal coupling $\xi$ on the probability of inflation $\mathcal{P}$ for the string-inspired fractional monomial potential  $V(\phi)=\lambda\phi^{1/3}$, with $M_{Pl}^{-11/3}\lambda \simeq 1.81\times10^{-10}$ inferred from the scalar power-spectrum amplitude  $\mathcal{P}_{s}=2.141 \times 10^{-9}$ \cite{Planck:2018jri}. The left panel shows the surface $\mathcal{P}(N,\xi)$ and the right panel shows the corresponding contour map. We fix the inverse-volume eigenvalue parameter to $l=3/4$ and the initial value to $q_i=100$.}
    \label{fig:1_3_xi}
\end{figure}

Following the same measure-theoretic approach, we calculate the probability of sufficient inflation for the fractional monomial potential $V(\phi) \propto \phi^{1/3}$. FIG.~\ref{fig:1_3_xi} maps the probability across the $(N, \xi)$ plane, utilizing a 3D surface plot on the left and a contour projection on the right. For this evaluation, $\lambda$ is fixed to satisfy the scalar power spectrum normalization. In stark contrast to the quartic case, the phase-space volume of favorable initial conditions for this fractional potential actually decreases as the non-minimal coupling grows. Although the axes are scaled to show a wide theoretical landscape spanning $\xi \in (0,1)$ up to $N=70$ ($M_{Pl}^2=1$), the actual physical regime that simultaneously yields a high probability of inflation and remains consistent with precision cosmological data occupies an extremely narrow strip of the configured parameter space.

\subsubsection{\texorpdfstring{$V(\phi)=\lambda \phi^{2/3}$}{V(phi) = lambda phi\textasciicircum(2/3)}}
Under the slow-roll approximation, we obtain
\begin{equation}
 \frac{d\phi}{dN}=\frac{2 \left(5 \xi  \phi^2-1\right) D_l^{n+1}}{3 \kappa  \phi}.
\end{equation}
So, the number of $e$-folds can be computed as 
\begin{equation}
     N =
    \frac{3 \kappa  \log \left(5 \xi  \phi ^2-1\right) D_l^{-(n+1)}}{20 \xi }\bigg|_{\phi=\phi_{i}}^{\phi=\phi_{f}}.
\end{equation}
Hence, by using \eqref{eq:probability_final_exp}, we get the Probability

\begin{equation}
    \begin{split}
        \mathcal{P}
        &\propto 2 a_{\text{Pl}}^3 q^{3/2}\sqrt{\frac{\lambda}{ 3\kappa}} \left[\sqrt{ \left(\xi  \phi _{i}^2+1\right)  \phi^{2/3}_{i} D_l(q_{i})^m}-\sqrt{\left(\xi  \phi_{f}^2+1\right)  \phi^{2/3}_{f}}\right].
    \end{split}
\end{equation}

\begin{figure}[H] 
    \centering
    \begin{minipage}{0.5\linewidth}
        \centering
        \includegraphics[width=0.8\textwidth]{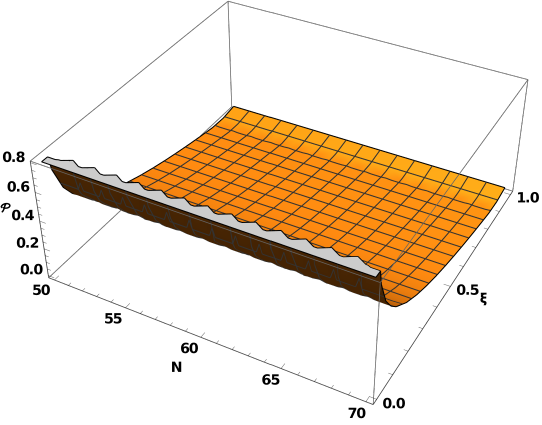}
    \end{minipage}\hfill
    \begin{minipage}{0.5\linewidth}
        \centering
        \includegraphics[width=0.8\textwidth]{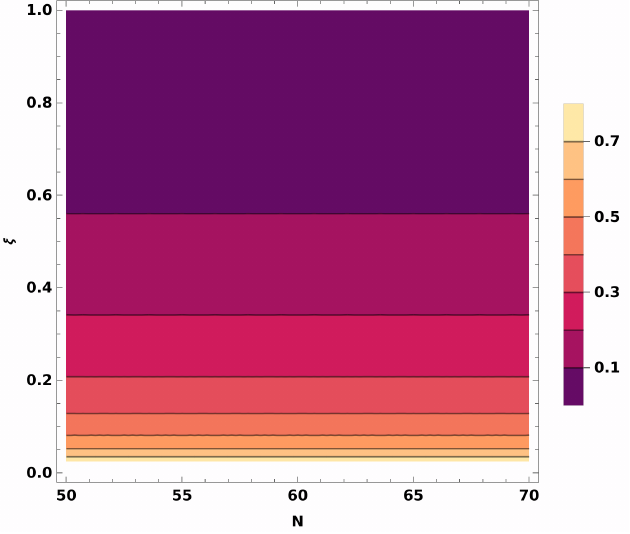}
    \end{minipage}
    \caption{Effect of the non-minimal coupling $\xi$ on the probability of inflation $\mathcal{P}$ for the string-inspired fractional monomial potential $V(\phi)=\lambda\phi^{2/3}$, with $M_{Pl}^{-10/3}\lambda \simeq 7.1\times10^{-10}$ inferred from the scalar power-spectrum amplitude   $\mathcal{P}_{s}=2.141\times 10^{-9}$ \cite{Planck:2018jri}. The left panel shows the surface $\mathcal{P}(N,\xi)$ and the right panel shows the corresponding contour map. We fix the inverse-volume eigenvalue parameter to $l=3/4$ and the initial value to $q_i=100$.}
    \label{fig:2_3_xi}
\end{figure}

Finally, we present the phase-space probability distributions for the $V(\phi) \propto \phi^{2/3}$ potential. As in the preceding analyses, the 3D and contour plots in FIG.~\ref{fig:2_3_xi} demonstrate how the non-minimal coupling $\xi$ and the total number of e-folds $N$ influence the measure of favorable initial conditions, assuming a fixed $\lambda$ calibrated by the primordial power spectrum. Similar to the $p=1/3$ scenario, we observe an inverse relationship between the coupling strength and the probability of inflation, with larger values of $\xi$ suppressing the likelihood of a sufficiently long slow-roll phase. We again plot this distribution over the wide macroscopic intervals $\xi \in (0,1)$ and $N \in (0,70)$ (with $M_{Pl}^2=1$) to clearly display the overall functional behavior of the Liouville measure. Nevertheless, one must keep in mind that the parameter configurations capable of surviving the stringent observational cuts from Planck and ACT represent only a very small fraction of the total parameter space visualized here.

The results summarized in TABLE~\ref{parameter_table} illustrate how inverse-volume corrections map onto the precision constraints of modern cosmology across three distinct potentials: $V(\phi) \propto \phi^{4}$, $\phi^{1/3}$, and $\phi^{2/3}$. By tracking the slow-roll parameters at $N=60, 65,$ and $70$ e-folds, we observe that the non-minimal coupling $\xi$ is the critical parameter governing observational viability. For the Higgs-like quartic potential, standard LQC dynamics yield an unacceptably large tensor-to-scalar ratio; however, values of $\xi \approx 0.04$ successfully suppress $r$ and tune the scalar tilt $n_s$ into excellent $1\sigma$ agreement with the combined Planck and ACT DR6 data. In contrast, the string-inspired fractional monomial potentials naturally predict lower tensor contributions, thus requiring much smaller non-minimal couplings ($\xi \sim 0.001$) to achieve the optimal $(n_s, r)$ and $(n_s, \alpha_s)$ fit. Across all three models, varying $\xi$ effectively acts to tilt the primordial spectrum and modulate its running, establishing a very narrow phenomenological window within which LQC-corrected slow-roll trajectories accurately reproduce the observed universe. The optimal value of $\xi$ for each potential is selected as the one whose predictions lie closest to the $1\sigma$ confidence region simultaneously in both the $(n_s, r)$ and $(n_s, \alpha_s)$ planes, and for both dataset combinations P-LB-BK18 and P-ACT-LB-BK18. This criterion is indicated in Table~\ref{parameter_table} by the highlighted $\xi$ values. Regarding the probability measure, it provides information about the Hamiltonian flow in the effective phase space and therefore characterizes the mathematical stability of the dynamical system. Concretely, it quantifies the fraction of post-bounce trajectories that converge to a sufficiently long slow-roll phase, and its enhancement with increasing (or decreasing) $\xi$ signals that the dynamics itself preferentially selects larger (or lower) values of the non-minimal coupling for Higgs potential (or monomial potentials). However, the physically admissible range of $\xi$ is ultimately restricted by the observational constraints on $(n_s, r, \alpha_s)$: the probability measure identifies the attractor structure of the flow, while the CMB data select the physically viable window within it. We also note that, in the presence of holonomy corrections, this same measure-theoretic framework can be used to assign a probability to the effective number of e-folds elapsed between the quantum bounce and the end of inflation, providing a natural bridge between the pre-inflationary LQC dynamics and the onset of slow-roll.

\begin{table}[H]
\caption{\label{parameter_table} Predicted inflationary observables ($n_s$, $r$, $\alpha_s$) and self-coupling $\lambda$ for the $\phi^4$, $\phi^{1/3}$, and $\phi^{2/3}$ potentials with LQC inverse-volume corrections. Results are evaluated at $N=60, 65,$ and $70$ e-folds for varying non-minimal coupling $\xi$. Colors denote consistency with Planck 2018 and ACT DR6 constraints in the $(n_s, r)$ and $(n_s, \alpha_s)$ planes: \colorbox{green}{\phantom{X}} ($\leq 1\sigma$), \colorbox{yellow}{\phantom{X}} ($\leq 2\sigma$), and \colorbox{red}{\phantom{X}} ($> 2\sigma$). Highlighted $\xi$ values indicate the optimal fit for each potential.
 }
\centering
\renewcommand{\arraystretch}{1.2}
\begin{tabular}{|c|c|c|c|c|c|c|c|c|c|c|}
\hline
\multirow{2}{*}{$\mathbf{V(\phi)}$} &\multirow{2}{*}{$\mathbf{\xi}$} &\multirow{2}{*}{$\mathbf{N}$} &\multirow{2}{*}{$\mathbf{n_s}$} &\multirow{2}{*}{$\mathbf{r}$} &\multirow{2}{*}{$\mathbf{\alpha_s}$} &\multirow{2}{*}{$\mathbf{\lambda}$} &\multicolumn{2}{|c|}{$\mathbf{P-LB-BK18}$} &\multicolumn{2}{|c|}{$\mathbf{P-ACT-LB-BK18}$}  \\ \cline{8-11}
& & & & & &  &$\mathbf{n_s-r}$ &$\mathbf{n_s-\alpha_s}$ &$\mathbf{n_s-r}$ &$\mathbf{n_s-\alpha_s}$ \\ \hline

\multirow{12}{*}{$\mathbf{\lambda\phi^{4}}$} &\multirow{3}{*}{$0.01$} &$60$ &$0.964$ &$0.045$ &$0.001$ &$2.06\times 10^{-13}$ &\cellcolor{red} &\cellcolor{green} &\cellcolor{red} &\cellcolor{red} \\ \cline{3-11}

 & &$65$ &$0.967$ &$0.039$ &$0.001$  &$1.73\times 10^{-13}$ &\cellcolor{red} &\cellcolor{green} &\cellcolor{red} &\cellcolor{yellow} \\ \cline{3-11}

  & &$70$ &$0.97$ &$0.034$ &$0$ &$1.48\times 10^{-13}$ &\cellcolor{yellow} &\cellcolor{yellow} &\cellcolor{red} &\cellcolor{yellow} \\ \cline{2-11}

  &\multirow{3}{*}{\fcolorbox{green}{white}{$0.04$}} &$60$ &$0.966$ &$0.013$ &$0.001$ &$7.19\times 10^{-13}$ &\cellcolor{green} &\cellcolor{green} &\cellcolor{yellow} &\cellcolor{red}  \\ \cline{3-11}

 & &$65$ &$0.969$ &$0.011$ &$0$ &$6.12\times 10^{-13}$&\cellcolor{green} &\cellcolor{green} &\cellcolor{yellow} &\cellcolor{yellow} \\ \cline{3-11}

  & &$70$ &$0.971$ &$0.01$ &$0$ &$5.27\times 10^{-13}$ &\cellcolor{green} &\cellcolor{yellow} &\cellcolor{green} &\cellcolor{green} \\ \cline{2-11}

  &\multirow{3}{*}{$0.07$} &$60$ &$0.967$ &$0.007$ &$0.001$ &$1.23\times 10^{-12}$ &\cellcolor{green} &\cellcolor{green} &\cellcolor{yellow} &\cellcolor{yellow} \\ \cline{3-11}

 & &$65$ &$0.969$ &$0.006$ &$0$ &$1.05\times 10^{-12}$ &\cellcolor{green} &\cellcolor{green} &\cellcolor{yellow} &\cellcolor{yellow} \\ \cline{3-11}

  & &$70$ &$0.971$ &$0.006$ &$0$ &$9.06\times 10^{-13}$ &\cellcolor{green} &\cellcolor{yellow} &\cellcolor{green} &\cellcolor{green} \\ \hline

\multirow{12}{*}{$\mathbf{\lambda\phi^{1/3}}$} &\multirow{3}{*}{$0.0005$} &$60$ &$0.975$ &$0.016$ &$0$ &$2.18\times 10^{-10}$ &\cellcolor{yellow} &\cellcolor{red} &\cellcolor{green} &\cellcolor{green} \\ \cline{3-11}

 & &$65$ &$0.977$ &$0.014$ &$0$ &$2.51\times 10^{-10}$ &\cellcolor{red} &\cellcolor{red} &\cellcolor{green} &\cellcolor{green} \\ \cline{3-11}

  & &$70$ &$0.978$ &$0.013$ &$0$ &$2.25\times 10^{-10}$ &\cellcolor{red} &\cellcolor{red} &\cellcolor{green} &\cellcolor{yellow} \\ \cline{2-11}

  &\multirow{3}{*}{\fcolorbox{green}{white}{$0.001$}} &$60$ &$0.97$ &$0.011$ &$0$ &$2.08\times 10^{-10}$ &\cellcolor{green} &\cellcolor{green} &\cellcolor{green} &\cellcolor{yellow} \\ \cline{3-11}

 & &$65$ &$0.972$ &$0.01$ &$0$ &$1.80\times 10^{-10}$ &\cellcolor{green} &\cellcolor{yellow} &\cellcolor{green} &\cellcolor{green} \\ \cline{3-11}

  & &$70$ &$0.973$ &$0.008$ &$0$ &$1.57\times 10^{-10}$ &\cellcolor{yellow} &\cellcolor{yellow} &\cellcolor{green} &\cellcolor{green} \\ \cline{2-11}

  &\multirow{3}{*}{$0.0015$} &$60$ &$0.965$ &$0.008$ &$0$ &$1.52\times 10^{-10}$  &\cellcolor{green} &\cellcolor{green} &\cellcolor{red} &\cellcolor{red} \\ \cline{3-11}

 & &$65$ &$0.966$ &$0.007$ &$0$ &$1.28\times 10^{-10}$ &\cellcolor{green} &\cellcolor{green} &\cellcolor{yellow} &\cellcolor{red} \\ \cline{3-11}

  & &$70$ &$0.967$ &$0.006$ &$0$ &$1.09\times 10^{-10}$ &\cellcolor{green} &\cellcolor{green} &\cellcolor{yellow} &\cellcolor{yellow} \\  \hline

  \multirow{12}{*}{$\mathbf{\lambda\phi^{2/3}}$} &\multirow{3}{*}{$0.0005$} &$60$ &$0.974$ &$0.032$ &$0$ &$7.96\times 10^{-10}$ &\cellcolor{red} &\cellcolor{yellow} &\cellcolor{green} &\cellcolor{green} \\ \cline{3-11}

 & &$65$ &$0.975$ &$0.028$ &$0$ &$7.15\times 10^{-10}$ &\cellcolor{red} &\cellcolor{red} &\cellcolor{green} &\cellcolor{green} \\ \cline{3-11}

  & &$70$ &$0.977$ &$0.026$ &$0$ &$6.47\times 10^{-10}$ &\cellcolor{red} &\cellcolor{red} &\cellcolor{green} &\cellcolor{yellow} \\ \cline{2-11}

  &\multirow{3}{*}{\fcolorbox{green}{white}{$0.001$}} &$60$ &$0.969$ &$0.023$ &$0$ &$7.87\times 10^{-10}$ &\cellcolor{green} &\cellcolor{green} &\cellcolor{yellow} &\cellcolor{yellow} \\ \cline{3-11}

 & &$65$ &$0.971$ &$0.02$ &$0$ &$7.07\times 10^{-10}$ &\cellcolor{green} &\cellcolor{yellow} &\cellcolor{green} &\cellcolor{yellow} \\ \cline{3-11}

  & &$70$ &$0.972$ &$0.017$ &$0$ &$6.41\times 10^{-10}$ &\cellcolor{yellow} &\cellcolor{yellow} &\cellcolor{green} &\cellcolor{green} \\ \cline{2-11}

  &\multirow{3}{*}{$0.0015$} &$60$ &$0.965$ &$0.016$ &$0$ &$7.88\times 10^{-10}$ &\cellcolor{green} &\cellcolor{green} &\cellcolor{yellow} &\cellcolor{red} \\ \cline{3-11}

 & &$65$ &$0.966$ &$0.014$ &$0$ &$7.10\times 10^{-10}$ &\cellcolor{green} &\cellcolor{green} &\cellcolor{yellow} &\cellcolor{red} \\ \cline{3-11}

  & &$70$ &$0.968$ &$0.012$ &$0$ &$6.46\times 10^{-10}$ &\cellcolor{green} &\cellcolor{green} &\cellcolor{yellow} &\cellcolor{yellow} \\ \hline
\end{tabular}

\end{table}


\section{Discussion and Conclusion}
\label{sec:Discussion_Conclusion}

In this work, we investigated slow-roll inflation driven by a scalar field non-minimally coupled to gravity within the effective framework of Loop Quantum Cosmology (LQC), incorporating inverse-volume corrections. For the two classes of potentials considered here, a Higgs-like quartic potential ($V \propto \phi^{4}$) and string-inspired fractional monomial potentials ($V \propto \phi^{p}$ with $p<1$), we computed the inflationary observables in the inverse-volume-corrected dynamics in the presence of non-minimal coupling and confronted the resulting predictions with the latest observational constraints. We then quantified the probability of achieving sufficient inflation by constructing the canonical Liouville measure on the effective phase space and evaluating the fraction of post-bounce trajectories that yield $N\ge N_\star$.

Our analysis shows that introducing a non-minimal coupling parameter $\xi$ significantly enlarges the phase space volume of initial conditions that lead to sufficient inflation, compared to the minimally coupled case. This attractor-like behavior saturates as the coupling strength increases, suggesting that non-minimal coupling renders inflation a generic outcome in these models within the LQC context. These results are particularly relevant in light of recent observational data. The Atacama Cosmology Telescope (ACT) DR6 and DESI analyses have indicated a preference for a scalar spectral index $n_s$ slightly higher than the Planck 2018 best-fit value, and the running $\alpha_s\equiv dn_s/d\ln k$ provides an additional discriminator among viable scenarios. We find that, within the inverse-volume-corrected effective LQC dynamics considered here, the non-minimally coupled fractional monomial potentials and the non-minimally coupled Higgs-like quartic model naturally fall in the preferred region of the $(n_s,r)$ plane, while the LQC bounce ensures the resolution of the initial singularity.

To connect our theoretical results on the probability of sufficient inflation with observations, we compare the parameter region where this probability is enhanced with the values of the non-minimal coupling $\xi$ that are consistent with current CMB constraints on $(n_s,r,\alpha_{s})$. For the Higgs-like quartic potential $V\propto\phi^{4}$, viable fits typically require $\xi=\mathcal{O}(10^{-2})$, with ACT-based combinations favoring values toward the upper end of this range. By contrast, for fractional monomial potentials $V\propto\phi^{p}$ with $p<1$ (e.g., $p=1/3$ and $p=2/3$), the preferred couplings are smaller, typically $\xi=\mathcal{O}(10^{-3})$, and can extend down to $\mathcal{O}(10^{-4})$ depending on the dataset and confidence level. Overall, the probability-enhanced region identified in our analysis overlaps with the order-of-magnitude values of $\xi$ that remain compatible with current observational bounds for both classes of non-minimally coupled potentials.

A central contribution of this work is to jointly assess the effects of non-minimal coupling and inverse-volume corrections on the slow-roll dynamics and the probabilistic realization of inflation, going beyond most existing analyses in the literature that consider these ingredients separately. In particular, we find that introducing the non-minimal coupling shifts the inverse-volume-corrected predictions toward values of $(n_{s}, r,\alpha_{s})$ that are more consistent with the ACT-preferred region, while simultaneously enlarging the phase-space measure of trajectories that yield sufficient inflation. Moreover, in the presence of inverse-volume corrections, the non-minimal coupling cannot, in general, be removed by field redefinitions \cite{Bojowald:2006hd,Han:2019mvj,Artymowski:2012is}, and therefore constitutes a physical ingredient of the effective dynamics, not merely a parametrization choice. A detailed quantitative comparison of the inverse-volume corrected probability against the standard classical GR expectation, isolating the net effect of the quantum geometric modifications from that of the non-minimal coupling, is provided in ~\ref{app_2_sec:Difference of Probability between the Inverse-Volume Corrected and the Classical}.

It is worth emphasizing the domain of validity of our effective treatment. We focused on inverse-volume corrections and worked in the regime where the effective description is reliable. Quantization ambiguities associated with these corrections, encoded in the parameters controlling inverse-volume effects and in operator-ordering choices, can affect quantitative predictions. In particular, such ambiguities may shift the detailed numerical values of the observables and the inferred probability, although the qualitative trend of an enhancement followed by saturation of the probability for $N\ge N_\star$ as $\xi$ increases remains stable across the representative ranges explored.

To further assess the robustness of our results, it is instructive to comment on the role of the parameters $m$, $n$, and $l$, which we have fixed throughout to $m=0$, $n=0$, and $l=3/4$. The choice $m=0$ and $n=0$ corresponds to the lowest-order operator-ordering prescription, i.e., the absence of any additional ordering ambiguity beyond the minimal one already encoded in $D_l(q)$. Increasing $m$ and $n$ to higher integer values introduces stronger inverse-volume weighting in the kinetic and potential terms of the effective Hamiltonian. Inspecting the analytic expressions for $n_s$ and $r$, one finds that higher values of $m$ and $n$ shift the scalar spectral index toward larger values (closer to $1$) and suppress the tensor-to-scalar ratio $r$. This trend is particularly interesting in light of the recent ACT DR6 data \cite{AtacamaCosmologyTelescope:2025blo, AtacamaCosmologyTelescope:2025nti}, which mildly favor a larger $n_s$ compared to Planck 2018 alone \cite{Planck:2018jri}, suggesting that non-minimal operator-ordering choices could further improve the agreement with current observations. The eigenvalue parameter $l \in (0,1)$ controls the shape of the correction function $D_l(q)$ and the rate of the quantum-to-classical transition. The standard choice $l=3/4$ is the most widely adopted in the LQC literature \cite{Bojowald:2002ny,Germani:2007rt} and yields results that are representative of the generic behavior. Varying $l$ would shift the quantitative predictions for the observables and the probability, but the overall conclusions regarding the observational viability of the models and the attractor-like probability enhancement driven by $\xi$ are expected to remain qualitatively stable across the range $l \in (0,1)$.

Our probability should be interpreted as a notion of typicality induced by the canonical Liouville measure on the effective constraint surface after gauge fixing. The enhancement with $\xi$ reflects a nontrivial re-weightting of the set of solutions. The non-minimal coupling changes the measure density on the transverse section used to evaluate the probability flux, thereby assigning greater weight to post-bounce trajectories that naturally converge to slow roll. This provides a dynamical explanation for why inflation becomes more generic in the non-minimally coupled case, beyond a mere shift of the slow-roll parameters.

To build further physical intuition, it is instructive to consider how the non-minimal coupling $\xi$ reshapes the effective potential landscape seen by the inflaton. In the Jordan frame, the coupling $\xi\phi^2 R$ makes the effective gravitational constant field-dependent, rendering the kinetic and potential terms in the equations of motion $\phi$-dependent 
in a correlated way. In the Einstein frame, this translates into a flattening of the effective potential at large field values, which broadens the basin of attraction of the slow-roll regime. As a consequence, a larger fraction of the post-bounce phase-space trajectories — which generically explore a wide range of field values and velocities after the quantum bounce — are naturally funneled into the slow-roll attractor, rather than overshooting it. This is precisely what the Liouville measure captures: the $\xi$-dependent terms in the measure element $B_q$ (Eq.~\eqref{eq:B_q}) assign greater flux weight to those initial conditions for which the post-bounce trajectory converges to a sustained slow-roll phase. As $\xi$ increases (or decreases), this funneling effect strengthens and eventually saturates, because beyond a certain coupling strength the potential is already sufficiently flat that further increases (or decreases) in $\xi$ do not significantly enlarge the basin of attraction. This saturation of the attractor-like enhancement at large (or small) $\xi$ is clearly visible in FIGS.~\ref{fig:4_xi}, \ref{fig:1_3_xi}, and \ref{fig:2_3_xi}, and provides a transparent physical picture of why non-minimal coupling renders inflation more generic within the LQC framework.

Several extensions follow naturally from this work. A first step is to study the combined impact of holonomy and inverse-volume corrections in the presence of non-minimal coupling, allowing the bounce and the onset of inflation to be treated within unified effective dynamics. It would also be valuable to refine the perturbation analysis by tracking the evolution of modes across the pre-inflationary transition region, where LQC corrections may leave characteristic imprints and where the mapping between $k$ and horizon crossing can be altered. Finally, incorporating reheating-consistent relations between $N_\star$ and post-inflationary expansion would enable a tighter confrontation with current and forthcoming data, particularly through constraints on the running and tensor modes.

Beyond the standard inflationary observables $(n_s, r, \alpha_s)$, the framework developed here suggests several broader connections worth highlighting. First, the inverse-volume corrections studied in this work are not unique to the inflationary sector: they arise generically in LQC as a consequence of the discrete quantum geometry of space, and their imprint on the primordial power spectrum could, in principle, leave observable signatures in the CMB at scales sensitive to the pre-inflationary dynamics, such as anomalies in the low-multipole power spectrum or modifications to the tensor power spectrum \cite{Ashtekar:2011ni, Agullo:2013dla}. Second, the non-minimal coupling $\xi\phi^2 R$ considered here is closely related to the broader class of scalar-tensor theories and Horndeski gravity \cite{Kobayashi:2019hrl}, where analogous attractor mechanisms have been identified in the classical setting. The LQC realization of this attractor, driven by the interplay between quantum geometry and the non-minimal coupling, provides a concrete ultraviolet completion of these classical scenarios and suggests that the preference for inflation may be a generic feature of quantum gravitational theories with non-minimal matter-geometry couplings. Third, the quantum bounce itself, which replaces the classical big-bang singularity in LQC, introduces a natural pre-inflationary phase whose dynamics could generate observable non-Gaussianities or superimposed oscillations in the primordial spectrum \cite{Ashtekar:2011ni}, providing a potential window onto quantum gravitational effects through forthcoming CMB and large-scale structure surveys such as the Simons Observatory \cite{Ade_2019}, CMB-S4 \cite{CMB-S4:2020lpa}, and LISA \cite{Ricciardone:2016ddg, LISACosmologyWorkingGroup:2025vdz}. Fourth, the measure-theoretic framework employed here to assess the typicality of inflation connects naturally to broader questions in quantum gravity about the initial state of the universe and the problem of time, suggesting that the probabilistic approach developed in this work could be extended to other quantum gravitational settings beyond LQC.

\section{Acknowledgments}
\label{sec:Acknowledgments}
We thank the COSMOGRAV-UTA group for its continuous support and stimulating research environment. R.R. is particularly grateful to Orestis Sarras for valuable discussions and Nitesh Kumar for providing the data and code used in this work. R.R. acknowledges financial support from the PhD fellowship of the UTA-ULS-UV consortium and FONDECYT Regular Grant No. 1220065. G. O. and J. S. acknowledge support from FONDECYT Regular Grant No. 1220065, Chile.

\section*{Data Availability}
\label{sec:Data Availability}
The datasets were derived from sources in the public domain: \linebreak
Planck (2018) - \url{https://esdcdoi.esac.esa.int/doi/html/data/astronomy/planck/Cosmology.html},\linebreak
ACT (DR6.02) - \url{https://lambda.gsfc.nasa.gov/product/act/act_dr6.02/act_dr6.02_chains_r_get.html}. \linebreak
All numerical implementations and data used in the computations presented 
in this work are publicly available in the GitHub repository 
\url{https://github.com/rudranilroy/arXiv-2603.04182}.

\appendix

\section{Measure}\label{app_1_sec:measure} 

\subsection{Construction}\label{app_1_subsec:Construction}
Before defining the measure, we should expect the measure to satisfy three basic conditions:
\begin{enumerate}
    \item It must be positive.
    \item It must be independent of the parameters chosen. The measure will be a property of the model, not the initial condition of the parameter.
    \item It must respect the symmetry of the space or the solution without introducing extra ad hoc structure. 
\end{enumerate}

Let us consider a system containing a finite number of variables and satisfying an ordinary differential equation with respect to the independent parameter ``time''. We consider  the evolution of the system as a flow in $N$-dimensional manifold $\theta$. Each trajectory of the flow represents a unique model of inflation. Now we can define a hypersurface $\Sigma$ intersecting the flow transversely in a $(N-1)$-surface $S$. We define a measure $\nu$ appropriate to the hypersurface to attach a weight $\mu(B)$ for each trajectory $B$ \cite{Gibbons:1986xk}:
\begin{equation}
    \mu (B)=\int_S \nu.
\end{equation}

Condition - 1 restricts $\mu(B)\geq 0$.

Condition - 2 removes the dependency of the measure on the chosen hypersurface. 

If we consider another hypersurface $\Sigma^\prime$ with intersection $S^\prime$, we will have the same measure
\begin{equation}
    \mu (B)=\int_{S^\prime} \nu.
\end{equation}

Now, regarding the differential flow, we may consider $S^\prime$ as some Lie transported version of $S$ along the tangent vector to the orbit, $V$. The invariance yields the condition
\begin{equation}
    \pounds_{fV} \ \nu=0,
\end{equation}
for an arbitrary function $f$.

Now we focus on the phase space dynamics. The system is described by the evolution in a $2n$-dimensional phase space $\Gamma_n$. The evolution of the system is given by the $2n$ first-order evolution equations
\begin{equation}
    \frac{dq^i}{dt}=\frac{\partial H(p,q)}{\partial p_i}, \qquad \frac{dp^i}{dt}=-\frac{\partial H(p,q)}{\partial q_i}.
\end{equation}
The integral curves form the Hamiltonian phase flow along the tangent direction
\begin{equation}
    X_H=\frac{\partial H(p,q)}{\partial p_i}\frac{\partial}{\partial q_i} - \frac{\partial H(p,q)}{\partial q_i}\frac{\partial}{\partial p_i}.
\end{equation}
As expected, the Hamiltonian remains invariant under the transportation along $X_H$
\begin{equation}
    \pounds_{X_H} H(p,q)=0.
\end{equation}

To work with more general approach, it is better to get rid of the coordinates. We constrain the dynamics to be confined within the hypersurface $\theta$ in $\Gamma_n$ with $H(p,q)=$constant. Within $\theta$, we chose a $(2n-2)$-dimensional hypersurface $\Gamma_{n-1}$, transverse to the Hamiltonian flow. To connect these spaces, we define the embedding of the manifolds through push-forward mapping $\iota_1:\theta \to \Gamma_n$, $\iota_2:\Gamma_{n-1}\to \theta$. As $\Gamma_n$ is even dimensional, we have a natural symplectic structure $\omega_n$. The closed non-degenerate differential 2-form $\omega_n$ is preserved by the Hamiltonian phase flow
\begin{equation}
    \pounds_{X_H} \omega_n=0.
\end{equation}

Following Darboux's theorem \cite{Arnold:1989who}, there must exist some local coordinates $(P_i,Q^i)$ such that $\omega_n$ takes the form
\begin{equation}
    \omega_n = dP_i \wedge dQ^i.
\end{equation}
If we introduce $P_1=H(p,q)$ and $Q^1=t$, we can write
\begin{equation}
    \omega_n = dH\wedge dt + \omega_{n-1},
\end{equation}
where $\omega_{n-1}$ is the symplectic structure in $\Gamma_{n-1}$.

Now we can define the pull-back of $\omega_n$ from $\Gamma_n$ to $\theta$
\begin{equation}
    \iota_1^* \omega_n = 0 \wedge dt + \omega_{n-1},
\end{equation}
and pull-back onto $\Gamma_{n-1}$ is
\begin{equation}
   \iota_2^* \iota_1^* \omega_n =  \omega_{n-1}.
\end{equation}

Hence, we have the natural Liouville measure on $\Gamma_{n}$ and $\Gamma_{n-1}$, respectively \cite{Gibbons:1986xk}
\begin{align} \label{eq_app_1:measure}
    \Omega_n &= (-1)^{n(n-1)/2} (\omega_n)^n\\
    \Omega_{n-1} &= (-1)^{(n-1)(n-2)/2} (\omega_{n-1})^{n-1}.
\end{align}
This measure already satisfies the conditions (1) and (3). We can check the condition (2) through the Cartan identity
\begin{equation}
    \pounds_{f X_H} (\iota_1^* \omega_n) = f X_H \ \lrcorner \ d(\iota_1^* \omega_n) + d(f X_H \ \lrcorner \ \iota_1^* \omega_n).
\end{equation}
where the symbol $\lrcorner$ denotes the interior product (or contraction). This operation inserts the vector field $f X_H$ into the first argument of the differential form, effectively mapping a $k$-form to a $(k-1)$-form. Since $\omega_n$ is a closed form, the exterior derivative $d$ and the pull-back commute, yielding $d(\iota_1^* \omega_n) = \iota_1^* d\omega_n=0$. We have $fX_H=f \ \partial/\partial Q^1$ in the Darboux coordinate. Thus, $f X_H \ \lrcorner \ \iota_1^* \omega_n=0$. Finally, we get the condition (2) as
\begin{equation}
    \pounds_{f X_H} (\iota_1^* \omega_n)=0.
\end{equation}
\subsection{Liouville Measures and Magnetic Flux}\label{app_1_subsec:Liouville Measures and Magnetic Flux}
 Though the natural measure $\Omega$ provides a valid measurement for the probability of inflation, it diverges due to the inclusion of infinitely many almost flat universes \cite{Gibbons:1986xk}. Gibbons and Turok \cite{Gibbons:2006pa} have defined an alternative measure based on the properties of a divergence-free field to describe the set of distinct dynamical trajectories or, equivalently, the set of classical initial conditions giving distinct histories. This new measure identifies all the nearly flat universes as the same, whose negligible differences cannot be detected from the observations.

We can identify the measure \eqref{eq_app_1:measure} with the flux of some divergence-free field, analogous to a ``magnetic field''. In general coordinates on phase space, we have the symplectic form \cite{Gibbons:2006pa}
\begin{equation}
    \omega_{\mu\nu} = - \omega_{\nu\mu},
\end{equation}
where $\mu, \nu = 1,2,3,\cdots,2n$, and $\det{[\omega]\neq 0}$. As $\omega$ is closed   ($d\omega=0$), we must have 
\begin{equation}
  d_{[\mu}\omega_{\nu\tau]}=0.
\end{equation}
The Hamiltonian equation is given by \cite{Gibbons:2006pa}
\begin{equation}
    V^\mu = \omega^{\mu\nu} \partial_\nu \mathcal{H},
\end{equation}
where $V^\mu\equiv (dx^\mu/dt)$ is the velocity in the phase space. We can write $\omega_{\mu\nu}V^\mu = \partial_\nu \mathcal{H}$, or equivalently 
\begin{equation}
     V^\mu \partial_\mu\mathcal{H}=0.
\end{equation}
Hence, the flow $V^\mu$ lies on the hypersurface $\mathcal{H}=$constant.

If we choose the coordinate $x^{2n}=\mathcal{H}$, the closure condition is restricted to the spatial indices
\begin{equation}
    d_{[a}\omega_{bc]}=0,
\end{equation}
and
\begin{equation}
    V^{2n}=0,
\end{equation}
with $a,b=1,2,3,\cdots,2n-1$.

As the Hamiltonian is conserved, i.e $\partial_t \mathcal{H}=0$, we must have
\begin{equation}
    V^a \omega_{ab}=0.
\end{equation}
As $\omega$ is closed, one can define a divergence-free field (analogous to a magnetic field)
\begin{equation}
    B_a\equiv\frac{1}{2}\epsilon_{abc}\omega_{bc},
\end{equation}
with
\begin{equation}
    \partial_a B_a =0.
\end{equation}
Hence,
\begin{equation}
    \epsilon_{abc}B_b V_c = 0,
\end{equation} that is to say  $\mathbf{V} \parallel \mathbf{B}$.

Fundamental considerations from the divergence theorem show that the flux through a fixed surface does not change under deformations that preserve its boundary \cite{Gibbons:2006pa}. Moreover, propagating the surface forward with the flow also leaves the flux unchanged.

For any phase space, we have a symplectic form \cite{Gibbons:1986xk}
\begin{equation} \label{eq:def_symplectic_form}
    \Omega = \sum_{i=1}^k \mathrm{d}P_i \wedge \mathrm{d}Q^i,
\end{equation}
where $Q_i$ and $P_i$ are the dynamical degrees of freedom and their conjugate momenta.

The phase space also contains a closed symplectic form
\begin{equation} \label{eq:def_closed_symplectic_form,ellis_multiverses_2004}
    \omega = \sum_{i=1}^{k-1} \mathrm{d}P_i \wedge \mathrm{d}Q^i,
\end{equation}
with the relation
\begin{equation}
    \Omega = \omega + \mathrm{d}\mathcal{H}\wedge \mathrm{d}t,  
\end{equation} and then 
\be
\omega = \Omega\Big|_{\mathcal{H}=0} \ .
\ee
Hence, these two symplectic forms are connected to each other through Hamiltonian constraint.

This construction can be extended to the effective Hamiltonian $\langle\widehat{\mathcal{H}_{LQ}}\rangle$ :
\begin{equation}
    \omega=\Omega\Big|_{\langle\widehat{\mathcal{H}_{LQ}}\rangle=0}
\end{equation}
From the symplectic form, we can define a divergence-free field \cite{Gibbons:2006pa}
\begin{equation} \label{eq:def_divergenceless_field}
    B_a=\frac{1}{2}\epsilon_{abc}\omega_{bc}.
\end{equation}
The field $\mathbf{B}$ defines the flow of trajectories across surfaces in the phase space. Due to the divergence-free nature of the field $\mathbf{B}$, we can define an associated vector potential $\mathbf{B}=d\mathbf{A}$. Hence, we can define a measure using Stokes' theorem \cite{Schiffrin:2012zf}
\begin{equation}\label{eq:def_measure}
    \mathcal{N}=\int \mathbf{B} \cdot \mathrm{d}\mathbf{S} = \oint \mathbf{A} \cdot \mathrm{d}\mathbf{l},
\end{equation}
where $\mathbf{S}$ is an open surface and  $\mathbf{l}=\partial\mathbf{S}$ is the boundary of $\mathbf{S}$. To avoid over-counting, the surface $\mathbf{S}$ is defined in such a way that the orbits cross through the surface only once.

The quantity $\mathcal{N}$ measures the number of trajectories crossing the surface $\mathbf{S}$ (or some equivalent surface) bounded by $\partial\mathbf{S}$.

\section{Difference of Probability between the Inverse-Volume Corrected and the Classical}
\label{app_2_sec:Difference of Probability between the Inverse-Volume Corrected and the Classical}

\begin{table}[H]
    \centering
     \caption{Change in Probability due to inverse-volume correction with $l=3/4$ and $q_i=100$. The values of self-coupling parameters $\lambda$ are fixed according to the scalar power spectrum.\vspace{1em}}
    \begin{tabular}{|c|c|}

        \hline

        $\mathbf{Potential}$ & $\mathbf{\lambda=Constant}$\\ 

        \hline & \\

        $V\propto\phi^4$ & 

        \adjustbox{valign=m}{\includegraphics[width=0.25\textwidth]{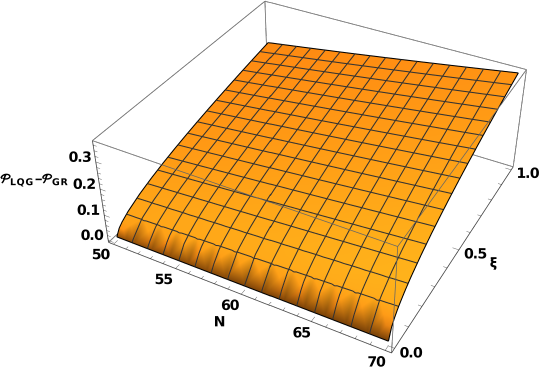}}  \adjustbox{valign=m}{\includegraphics[width=0.2\textwidth]{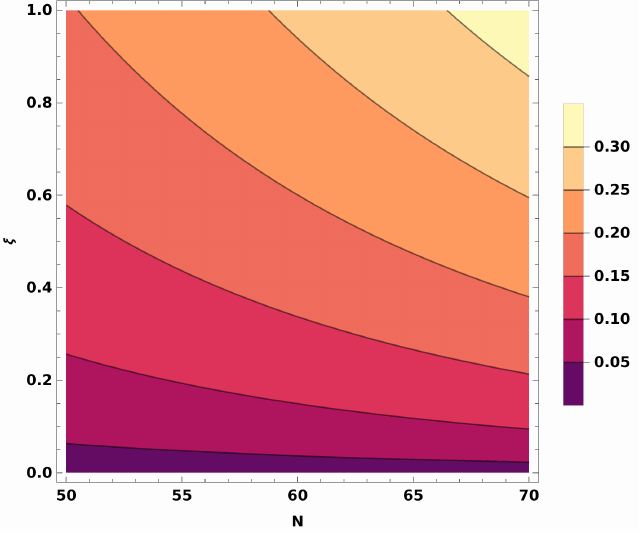}}  \\

        \hline

        $V\propto\phi^{1/3}$ & 

        \adjustbox{valign=m}{\includegraphics[width=0.25\textwidth]{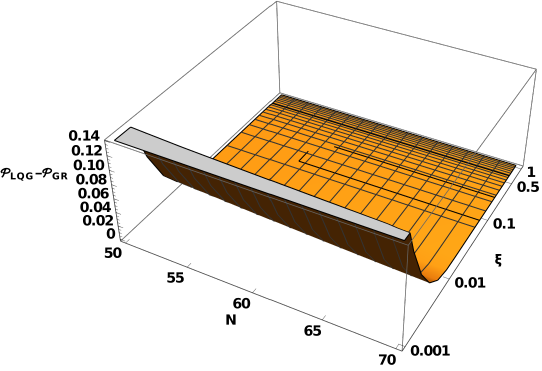}}  \adjustbox{valign=m}{\includegraphics[width=0.2\textwidth]{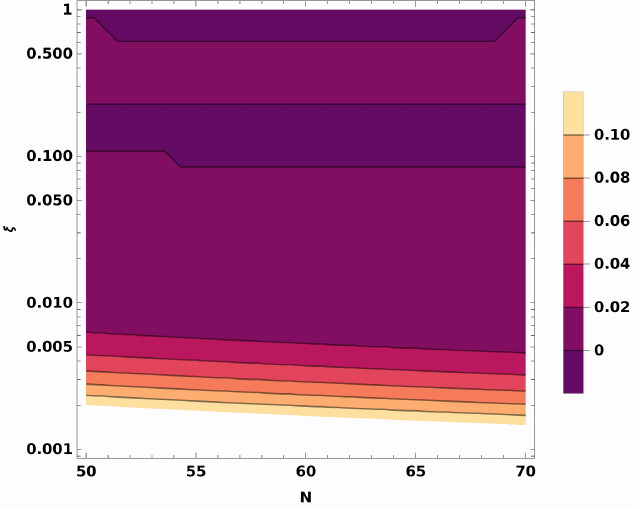}} \\

        \hline

        $V\propto\phi^{2/3}$ & 

        \adjustbox{valign=m}{\includegraphics[width=0.25\textwidth]{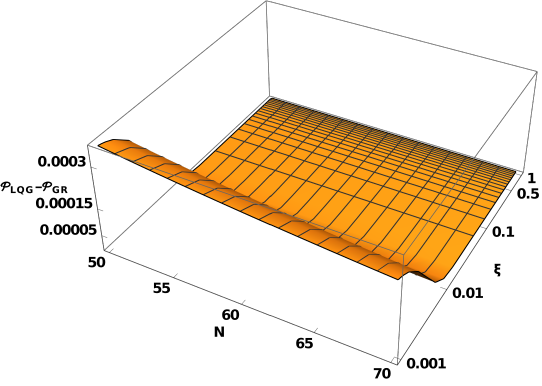}}  \adjustbox{valign=m}{\includegraphics[width=0.2\textwidth]{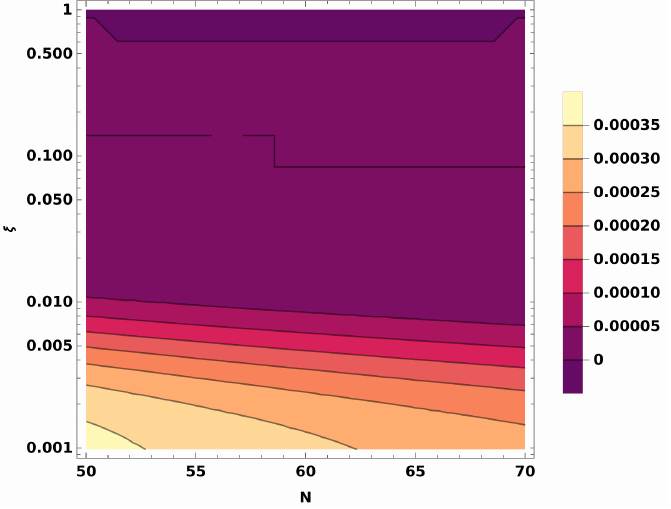}} \\

        \hline 

    \end{tabular}
    \label{fig:prob_diff}

\end{table}

To explicitly isolate the impact of the quantum geometric modifications, this appendix compares the probability of inflation derived from the effective non-minimally coupled LQC dynamics against the standard classical General Relativity (GR) expectation. To disentangle the two distinct physical ingredients of our framework, it is useful to compare their respective roles separately. The inverse-volume corrections, encoded in $D_l(q)$ and evaluated at horizon crossing ($q_i=100 \gg 1$), are perturbatively small at the level of the inflationary observables $n_s$, $r$, and $\alpha_s$, since $D_l(q_i) > 1$ but close to unity. Their primary impact is instead on the phase-space measure: through the factor $D_l(q_i)^m$ appearing in the Liouville measure element $B_q$ (Eq.~\eqref{eq:B_q}), the inverse-volume corrections directly re-weight the distribution of post-bounce trajectories, enhancing or suppressing the phase-space volume of favorable initial conditions relative to the classical GR case. The non-minimal coupling $\xi$, by contrast, affects both aspects simultaneously. It changes the Einstein-frame effective potential and, consequently, the slow-roll hierarchy, generally suppressing the tensor-to-scalar ratio $r$ through a reduction of the effective slope. The corresponding shift in the scalar spectral index $n_s$, however, is not universal. It depends on the potential through the $\xi$-induced changes in the slow-roll parameters $\epsilon_V$ and $\eta_V$, which encode the effective slope and curvature of the potential, respectively. Thus, in the Higgs-like case studied here, $n_s$ is shifted toward larger values, whereas in the monomial case, it is shifted toward smaller values. In addition, $\xi$ reshapes the Liouville measure through the $\xi$-dependent contributions to $B_q$, see Eq.~\eqref{eq:B_q}. The net result is that the non-minimal coupling provides the dominant driver of the attractor-like probability enhancement identified in this work, while the inverse-volume corrections provide an additional, subdominant but physically distinct, quantum geometric contribution. Table~\ref{fig:prob_diff} presents this comparison across all three studied potentials: $V(\phi) \propto \phi^4, \phi^{1/3}$, and $\phi^{2/3}$. For each model, we calculate the absolute difference between the inverse-volume corrected probability and the uncorrected classical probability, plotting the result as a function of the non-minimal coupling parameter $\xi$ and the number of e-folds $N$. In these evaluations, the scale parameter $\lambda$ is continuously fixed to satisfy the scalar power spectrum normalization, while the LQC parameters are set to $l=3/4$ and the initial state to $q_i=100$. The 3D surface plots and their corresponding contour maps clearly visualize the net topological effect of the loop quantum corrections, revealing exactly where in the parameter space the inverse-volume modifications enhance or suppress the phase-space volume of favorable initial conditions relative to the standard classical framework.

\bibliographystyle{elsarticle-harv} 
\bibliography{bio}






\end{document}